\documentclass[apj]{emulateapj}

\newcommand{\sm}{M_\odot}
\def\farcsec{\hbox{$.\!\!^{\prime\prime}$}}
\shorttitle{PTF and {\it GALEX} Light Curves of M31 Novae}
\shortauthors{Cao et al.}

\begin{document}
\title{Classical Novae in Andromeda: Light Curves from the
	Palomar Transient Factory and {\it GALEX}}
\author{Yi~Cao\altaffilmark{1,2}, Mansi~M.~Kasliwal\altaffilmark{2,3},
	James~D.~Neill\altaffilmark{2}, S.~R.~Kulkarni\altaffilmark{2},
        Yu-Qing~Lou\altaffilmark{1}, Sagi~Ben-Ami\altaffilmark{4}
	Joshua~S.~Bloom\altaffilmark{5},
	S.~Bradley~Cenko\altaffilmark{5}, Nicholas~M.~Law\altaffilmark{6},
        Peter~E.~Nugent\altaffilmark{5,7}, Eran~O.~Ofek\altaffilmark{8},
	Dovi~Poznanski\altaffilmark{9} \&\ Robert~M.~Quimby\altaffilmark{2,10}
        }

\altaffiltext{1}{Tsinghua Center for Astrophysics (THCA),
	Department of Physics, Tsinghua University, Beijing, 100084, China;
	ycao@astro.caltech.edu}
\altaffiltext{2}{California Institute of Technology, 1200 E California Blvd.,
	Pasadena, CA 91125, USA}
\altaffiltext{3}{Observatories of the Carnegie Institution for Science, 
        813 Santa Barbara St, Pasadena, CA, 91101, USA}
\altaffiltext{4}{Department of Particle Physics and Astrophysics, The 
        Weizmann Institute of Science, Rehovot 76100, Israel}
\altaffiltext{5}{Department of Astronomy, University of California,
	Berkeley, CA 94720-3411, USA}
\altaffiltext{6}{Dunlap Institute for Astronomy and Astrophysics,
	University of Toronto, 50 St. George Street, Toronto M5S 3H4,
	Ontario, Canada}
\altaffiltext{7}{Computational Cosmology Center,
	Lawrence Berkeley National Laboratory, 1 Cyclotron Road,
	Berkeley, CA 94720, USA}
\altaffiltext{8}{Benoziyo Center for Astrophysics, Faculty of Physics, 
        The Weizmann Institute of Science, Rehovot 76100, Israel}
\altaffiltext{9}{School of Physics and Astronomy, Tel-Aviv University, 
        Tel Aviv 69978, Israel}
\altaffiltext{10}{IPMU, University of Tokyo, Kashiwanoha 5-1-5, Kashiwa-shi, 
        Chiba, Japan}

\begin{abstract}

We present optical light curves of twenty-nine novae in M31 during the
2009 and 2010 observing seasons of the Palomar Transient Factory (PTF). The
dynamic and rapid cadences in PTF monitoring of M31, from one day to even ten
minutes, provide excellent temporal coverage of nova light curves, enabling
us to record the photometric evolution of M31 novae in unprecedented
detail.  We also detect eight of these novae in the near ultraviolet (UV)
band with the Galaxy Evolution Explorer ({\it GALEX}) satellite. Novae
M31N\,2009-10b and 2010-11a show prominent UV emission peaking a few days
prior to their optical maxima, possibly implying aspherical outbursts.
Additionally, our blue-shifted spectrum of the recent outburst of PT And
(M31N\,2010-12a) indicates that it is a recurrent nova in M31 and not a
dwarf nova in the Milky Way as was previously assumed.  Finally, we
systematically searched for novae in all confirmed globular clusters of M31
and found only M31N 2010-10f associated with Bol~126.  The specific nova rate 
in the
M31 globular cluster system is thus about one per year which is not enhanced
relative to the rate outside the globular cluster system.

\end{abstract}

\keywords{surveys -- nova, cataclysmic variables}

\section{INTRODUCTION}
Classical novae (CNe) are interesting to study for at least two reasons: 
first, they provide a laboratory to investigate thermonuclear runaways in 
semidegenerate conditions \citep[e.g., ][]{BE2008}; second, they have been 
proposed as progenitors of type Ia supernovae \citep[e.g., ][]{SSS1988}. 
Observations of Galactic novae are limited due to extinction
and distance determination. CNe are frequent but faint relative to supernovae.
Hence, people pursue novae in the nearest galaxies.

CNe in M31 (The
Andromeda Galaxy) have been observed for nearly a century
\citep{Hubble1929, Arp1956, Rosino1964, Rosino1973, CFN1987, SA1991,
TS1992, RJC1999, CLR2004, SI2001, DBK2004, DBK2006, KCK2010, SBD2011}.
The published nova rates for M31 range between 24~yr$^{-1}$ \citep{Arp1956} and
65~yr$^{-1}$ \citep{DBK2004}. 

The Palomar Transient Factory (PTF) \citep{LKD2009, RKL2009} acquired
nearly daily monitoring of M31 in the R band during the 2009 and 2010
observing seasons. It provides three advantages for the study of novae in M31:
first, the large field of view (FoV) of 7.2 square degrees covers the main
region of M31 in one snapshot and with a spatial resolution of $1\farcsec0$. Second,
PTF has fast and dynamic cadences from one day to as short as ten minutes.
This temporal coverage gives unprecedentedly detailed light curves.
Third, the limiting magnitude is deeper than 20\,mag (a little
worse near the center of M31 where the background is strong and varies
quickly).  Hence, given the mean distance modulus of 24.36 for M31 
\citep{VRJ2010}, 
unless some very faint novae are extincted heavily by local dust, 
PTF should see nearly all of the novae (weather-permitting) in M31 and
follow them up down to a magnitude of M$_R \sim-4.4$. 

These data
also afford the opportunity to explore known recurrent novae (RNe) in M31
and perhaps discover new ones. 
According to the catalog maintained by \citet{Pietsch2010}, 
about thirteen recurrent novae have been discovered in M31. 
RNe are possible progenitors of Type Ia 
supernovae, since their white dwarf mass is thought to be closer to the 
Chandrasekhar mass limit than ordinary CNe \citep{SSS1988}

Another issue related to novae that can be assessed with PTF archives is whether the nova
rate in the globular cluster (GC) system is enhanced relative to that outside the globular cluster system.
\citet{CTP1990} did a search for novae in fifty-four M31 GCs based on the H$\alpha$ survey of 
\citet{CFN1987} and \citet{CSF1990}. In a mean effective survey time of approximate two
years, they did not find any nova outbursts in any of the M31 GCs. 
\citet{TCS1992} reported another one-year search of over two hundred M31 GCs with H$\alpha$ emission
but did not find any nova outbursts, providing an upper limit on the rate of 
0.005~nova~yr$^{-1}$~GC$^{-1}$.  The only GC nova found in M31 is M31N\,2007-06b \citep{SQ2007}. 
\citet{HPH2009} then derived a rate of 1.1~nova~yr$^{-1}$ for the M31 GC system, which
is comparable to the rate (per unit mass) outside the globulars.  A systematic search for novae in M31 GCs using the 
PTF archives is presented here and also supported this conclusion. 

In a nova outburst, except for a short ``fireball'' phase where
radiation is given out by the thermal emission of the photosphere, 
UV photons are thought to originate at the surface of
the white dwarf (WD), deep within the nova envelope, while optical photons are 
released
near the surface of the envelope and result from absorption and re-emission
of relatively low ionization species such as Cr~II, Fe~II and other heavy
metals generated during the outburst.  Nova shells therefore serve as
passive photon converters whose spectra reflect the reprocessing of
incident UV light to longer wavebands.  Eventually, the expansion decreases
the temperature and dilutes absorbing species to finally reveal the UV
photons \citep{KH1994,GK1998}.  Therefore, to explore the UV light powering the nova emission, we
searched in parallel for any contemporaneous ultra-violet observations of
CNe in M31 with the {\it GALEX} satellite \citep{MFS2005}.  
Finally, we obtained near UV-optical light curves
for eight novae in M31.

This paper is arranged as follows. Section~\ref{obs} describes PTF and {\it
GALEX} observation and data reduction as well as data from other sources. 
Section~\ref{section_LC} presents light curves of classical novae and describes their
morphology. Section~\ref{section_gc} discusses novae in globular clusters. 
Section~\ref{section_RN} focuses on recurrent novae. Section~\ref{section_UV} discusses
ordinary and bizarre novae observed in UV band. Section~\ref{section_MMRD} revisits 
the MMRD relation with our novae sample. Finally we conclude this paper in Section~\ref{conclusion}. 

\section{OBSERVATION}
\label{obs}

\subsection{Nova Sample}

\begin{figure*}
\begin{center}
\includegraphics[width=0.8\textwidth]{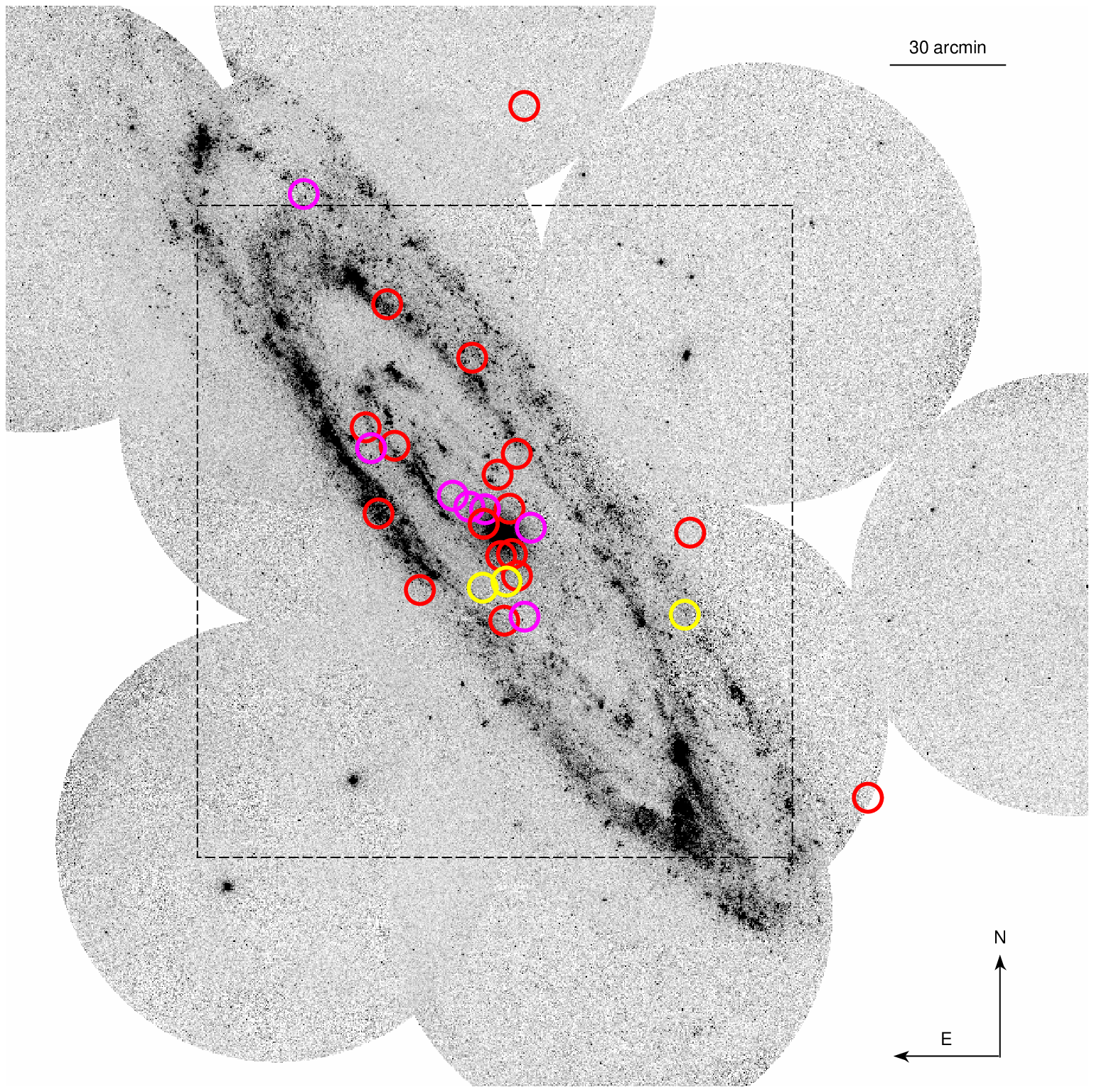}
\includegraphics[angle=0,width=0.6\textwidth]{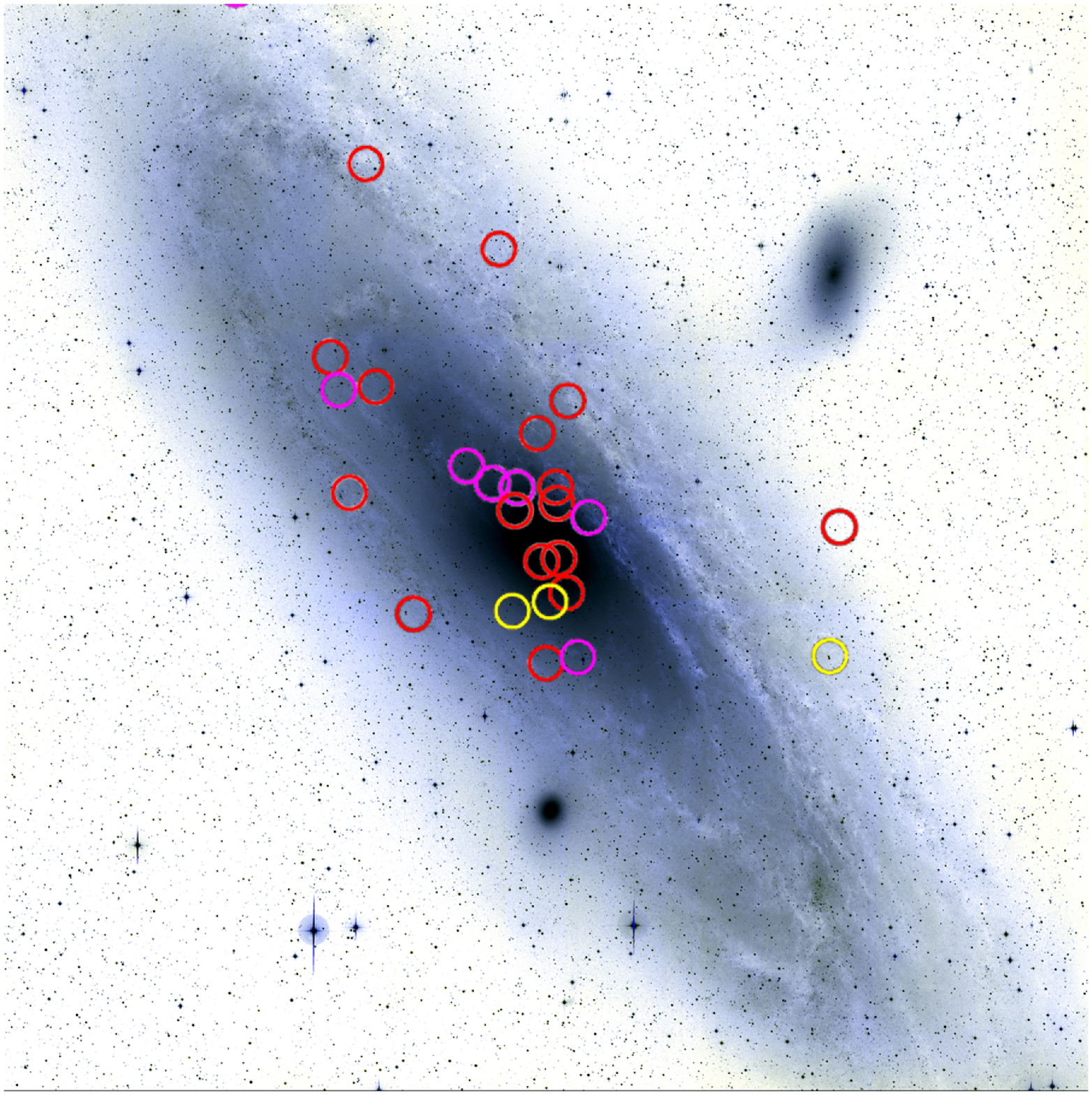}
\caption{The spatial distribution of the nova sample in M31. The upper panel
is an image of M31 taken by {\it GALEX} \citep{GBM2007}.
The bottom panel is a mosaic image of M31 from PTF (Credit: P.~Nugent).
The dashed box in the upper panel shows the physical size of the bottom one. 
The actual FoV is larger than the size of this image. 
The red circles
denote positions of novae only observed in PTF. RNe are
highlighted by the yellow circles.  The magenta circles denote novae
observed in both PTF and {\it GALEX}. 
\label{spatial_distribution}}
\end{center}
\end{figure*}

We examined all novae as reported by
\citet{Pietsch2010}\footnote{http://www.mpe.mpg.de/$\sim$m31novae/opt/m31/index.php. This 
site maintains a comprehensive catalog of novae in M31.}
in the PTF image archives and obtained the light curves of twenty-nine novae.
Among them, eight novae also have UV light curves from {\it GALEX}.  Table
\ref{obs_sum} gives main parameters of PTF and {\it GALEX} observations.  Table
\ref{Nova_List} describes the nova sample and Figure
\ref{spatial_distribution} illustrates their spatial distribution.

\begin{deluxetable}{ccc}
  \tablecaption{Observation summary \label{obs_sum}}
  \tablehead{\colhead{} & \colhead{PTF} & \colhead{{\it GALEX}} }
  \startdata
  Filter & Mould R & NUV \\
  $\lambda_{\rm effect}$ (\AA) & 6581 & 2316 \\
  $\Delta\lambda$ (\AA) & 1251 & 1060 \\
  $t_{\rm expose}$ (s) & 60 & 740-1703 \\
  FoV (deg$^2$) & 7.2 & 1.2 \\
  Cadence & 10min to days & 1-2 days \\
  Epoch & 726 & 51 (the center field)\\
             &        & 22-27 (outskirt fields)
  \enddata
\end{deluxetable}

\begin{figure}
\plotone{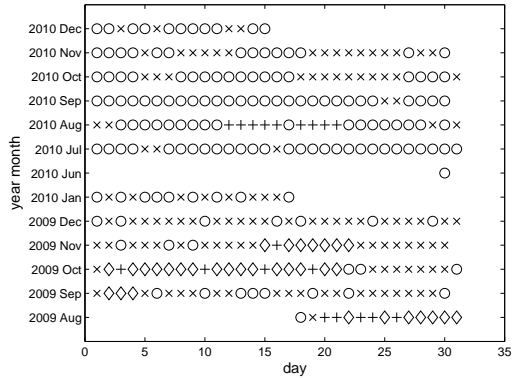}
\caption{The temporal coverage of both PTF and {\it GALEX}.  A plus sign
indicates an epoch observed by both PTF and {\it GALEX} while a cross indicates
epochs observed by
neither.  PTF-only observations are denoted with circles.  {\it GALEX}-only
observations are represented by diamonds.  Statistically, during the first
observing season, PTF observed M31 for $20\%$ of 156 total nights, {\it
GALEX} observed for $21\%$ nights and the two overlapped on $6\%$
nights.  In the second season, PTF observed for $75\%$ of 169 nights and
{\it GALEX} overlapped with PTF on nine nights.
\label{Temp_Cover}}
\end{figure}

\begin{center}
\begin{deluxetable*}{cccccccc}
    \tablecaption{Sample of Novae\label{Nova_List}}
    \tablehead{
        \colhead{} & \colhead{$\alpha$} & \colhead{$\delta$} &
        \colhead{Discovery Date} & \colhead{Magnitude} & \colhead{PTF First Detection} & \colhead{Spectral} & \colhead{} \\
        \colhead{Nova Name} & \colhead{(J2000)} & \colhead{(J2000)} &
        \colhead{by Others} & \colhead{at Discovery} & \colhead{Date}
        & \colhead{Type} & \colhead{Reference}
    }
    \startdata
    M31N\,2009-08b & $00^h44^m09^s.87$ & $+41^\circ48'50\farcsec8$ & 2009/08/09.780 & 17.2(U)\tablenotemark{1} & 2009/08/18.423 & ... & 1 \\
    M31N\,2009-08e & $00^h42^m36^s.23$ & $+41^\circ18'01\farcsec6$ & 2009/08/25.896 & 17.9(R) & 2009/08/26.400 & FeII & 2 \\
    M31N\,2009-09a & $00^h42^m26^s.08$ & $+41^\circ04'01\farcsec0$ & 2009/09/02.078 & 17.1(U) & 2009/08/18.454 & FeII & 3 \\
    M31N\,2009-10a & $00^h45^m14^s.04$ & $+42^\circ04'38\farcsec8$ & 2009/10/03.619 & 17.1(U) & 2009/09/30.494 & FeII & 4 \\
    M31N\,2009-10b & $00^h42^m20^s.77$ & $+41^\circ16'44\farcsec5$ & 2009/10/11.414 & 18.8(R) & 2009/10/10.376 & FeII & 5 \\
    \cr
    M31N\,2009-11a & $00^h43^m04^s.76$ & $+41^\circ41'08\farcsec2$ & 2009/11/03.559 & 17.8(U) & 2009/10/31.321 & FeII & 6 \\
    M31N\,2009-11b\tablenotemark{2} & $00^h42^m39^s.61$ & $+41^\circ09'03\farcsec2$ & 2009/11/06.523 & 18.6(R) & 2009/11/07.199 & ... & 7 \\
    M31N\,2009-11d & $00^h44^m16^s.85$ & $+41^\circ18'53\farcsec6$ & 2009/11/19.194 & 18.1(R) & 2009/12/01.201 & FeII & 8 \\
    M31N\,2009-11e & $00^h42^m35^s.31$ & $+41^\circ12'59\farcsec1$ & 2009/11/21.138 & 18.8(Swift uvw2) & 2009/12/01.201 & FeII & 9 \\
    M31N\,2009-12a & $00^h40^m19^s.40$ & $+41^\circ15'47\farcsec6$ & 2009/12/22.488 & 16.6(U) & 2009/12/10.109 & ... & 10 \\
    \cr
    M31N\,2010-01a\tablenotemark{3} & $00^h42^m56^s.74$ & $+41^\circ17'21\farcsec0$ & 2010/01/11.13 & 17.6(R) & 2010/01/11.130 & FeII & 11 \\
    M31N\,2010-06a & $00^h43^m07^s.56$ & $+41^\circ19'49\farcsec0$ & 2010/06/28.014 & 18.1(R) & 2010/06/30.397 & FeII & 12 \\
    M31N\,2010-06b & $00^h44^m22^s.46$ & $+41^\circ28'14\farcsec5$ & 2010/06/28.014 & 19.1(R) & 2010/06/30.353 & FeII & 13 \\
    M31N\,2010-06c & $00^h44^m04^s.48$ & $+41^\circ28'34\farcsec2$ & 2010/06/26.084 & 17.8(R) & 2010/06/30.353 & ... & 14 \\
    M31N\,2010-06d & $00^h42^m55^s.61$ & $+41^\circ19'26\farcsec0$ & 2010/06/24.02 & 19.5(Swift uvw1) & 2010/07/04.397 & FeII & 15 \\
    \cr
    M31N\,2010-07a & $00^h43^m20^s.11$ & $+41^\circ21'23\farcsec7$ & 2010/07/07.036 & 20.6(Swift uvw1) & 2010/07/07.328 & FeII & 16 \\
    M31N\,2010-09a & $00^h42^m23^s.32$ & $+42^\circ17'08\farcsec6$ & 2010/09/04.552 & 17.2(U) & 2010/09/01.372 & FeII & 17 \\
    M31N\,2010-09b & $00^h43^m45^s.53$ & $+41^\circ07'54\farcsec7$ & 2010/09/30.412 & 17.7(R) & 2010/09/30.230 & FeII & 18 \\
    M31N\,2010-09c & $00^h38^m09^s.06$ & $+40^\circ37'25\farcsec9$ & 2010/09/03.30 & 18.7(U) & 2010/08/13.235 & FeII & 19 \\
    M31N\,2010-10a & $00^h42^m45^s.84$ & $+41^\circ24'22\farcsec2$ & 2010/10/05.551 & 17.6(R) & 2010/10/08.163 & FeII & 20 \\
    \cr
    M31N\,2010-10b & $00^h42^m41^s.51$ & $+41^\circ03'27\farcsec3$ & 2010/08/19.055 & 18.9(R) & 2010/10/08.163 & FeII & 21 \\
    M31N\,2010-10c & $00^h44^m26^s.56$ & $+41^\circ31'13\farcsec8$ & 2010/10/13.557 & 17.8(U) & 2010/10/13.113 & FeII & 22 \\
    M31N\,2010-10d & $00^h42^m36^s.91$ & $+41^\circ19'29\farcsec6$ & 2010/10/29.478 & 17.8(U) & 2010/10/29.213 & FeII & 23 \\
    M31N\,2010-10e\tablenotemark{4} & $00^h42^m57^s.76$ & $+41^\circ08'12\farcsec3$ &  2010/10/31.727 & 18.1(R) & 2010/11/01.158 & He/N & 24 \\
    M31N\,2010-10f\tablenotemark{5} & $00^h42^m43^s.58$ & $+41^\circ12'42\farcsec6$ & 2010/10/11.10 & ... & 2010/10/11.110 & ... & 25 \\
    \cr
    M31N\,2010-11a & $00^h42^m31^s.60$ & $+41^\circ09'51\farcsec5$ & 2010/11/23.05 & 18.2(Swift uvw1) & 2010/12/01.250 & He/N & 26 \\
    M31N\,2010-12a\tablenotemark{6} & $00^h40^m24^s.37$ & $+41^\circ04'03\farcsec5$ & 2010/10/01.15 & 16.7(U) & 2010/12/01.270 & FeII & 27 \\
    M31N\,2010-12b & $00^h42^m31^s.08$ & $+41^\circ27'20\farcsec3$ & 2010/12/11.095 & 16.7(U) & 2010/12/10.115 & ... & 28 \\
    M31N\,2010-12c\tablenotemark{3} & $00^h42^m56^s.67$ & $+41^\circ17'21\farcsec2$ & 2010/12/11.095 &17.2(U) & 2010/12/15.093 & He/N & 29
    \enddata

    \tablerefs{(1) \citealp{ATel2166,ATel2176}; (2) \citealp{ATel2176,ATel2274,ATel2213};
    (3) \citealp{ATel2189,ATel2205}; (4) \citealp{ATel2239};
    (5) \citealp{CBET1967,CBET1971p1,CBET1971p2,CBET1971p3,CBET1971p4,CBET1971p5,CBET1973,CBET1980p1,CBET1980p2,ATel2248,ATel2251,ATel2274};
    (6) \citealp{CBET2003p1,CBET2003p2,CBET2062}; 
    (7) \citealp{CBET2015p1,CBET2015p2,CBET2015p3,ATel2286,ATel2290};
    (8) \citealp{CBET2058p1,CBET2058p2,CBET2058p3,ATel2304};
    (9) \citealp{CBET2061p1,CBET2061p2,CBET2061p3,CBET2061p4,CBET2061p5,ATel2308};
    (10) \citealp{CBET2100}; 
    (11) \citealp{CBET2124,CBET2127,CBET2136,CBET2187p1,CBET2187p2,CBET2187p3,Atel2383,ATel2435};
    (12) \citealp{CBET2341p1,CBET2341p2,CBET2391p1,CBET2391p2,CBET2391p3,ATel2787,ATel2964};
    (13) \citealp{CBET2342,CBET2391p1,CBET2391p2};
    (14) \citealp{CBET2343p1,CBET2343p2,CBET2343p3,CBET2391p1,CBET2391p3,ATel2697};
    (15) \citealp{CBET2347p1,CBET2347p2,ATel2713,ATel2787,ATel2789,ATel2964};
    (16) \citealp{CBET2358p1,CBET2358p2,CBET2391p1,CBET2391p2,ATel2727,ATel2787,ATel2964};
    (17) \citealp{CBET2448,ATel2834}; (18) \citealp{CBET2472,ATel2896,ATel2898,ATel2909};
    (19) \citealp{ATel2840,ATel2843,ATel2844}; 
    (20) \citealp{CBET2483,ATel2909,ATel2964}; (21) \citealp{CBET2487,ATel3039};
    (22) \citealp{CBET2500,ATel2949};
    (23) \citealp{CBET2516p2,CBET2516p3,CBET2516p4,CBET2571,ATel2987};
    (24) \citealp{CBET2573,ATel3001,ATel3006,ATel3038}
    (25) \citealp{ATel3013,ATel3019,ATel3074};
    (26) \citealp{ATel3061,ATel3066,ATel3068};
    (27) \citealp{CBET2574};
    (28) \citealp{CBET2582p1,CBET2582p2,CBET2582p3,CBET2582p4,CBET2582p5,CBET2582p6,CBET2582p7,ATel3076};
    (29) \citealp{CBET2594,CBET2597p1,CBET2597p2,CBET2610p1,CBET2610p2};
    }
    \tablenotetext{1}{The bracket following the magnitude indicates the observation band and U represents unfiltered observation.}
    \tablenotetext{2}{RN; 1997-11k=2001-12b=2009-11b.}
    \tablenotetext{3}{According to CBET 2610, 2010-01a and 2010-12c are two different objects within $0".8$.}
    \tablenotetext{4}{RN; 1963-09c=1968-09a=2001-07b=2010-10e.}
    \tablenotetext{5}{The nova 2010-10f is probably inside the Globular Cluster Bol~126.}    
    \tablenotetext{6}{Another recurrent burst of PT And.}
\end{deluxetable*}
\end{center}

\begin{center}
  \begin{deluxetable*}{cccc}
   \tablecaption{Novae observed with {\it GALEX} \label{galexobs}}
    \tablehead{
      \colhead{Name} & \colhead{Epochs} & \colhead{UT start} &
      \colhead{UT end}}
    \startdata
    M31N\,2009-09a & 63 & 2009/08/20.429 & 2009/10/15.707 \\
    M31N\,2009-10a & 13 & 2009/10/02.629 & 2009/10/21.457 \\
    M31N\,2009-10b & 22 & 2009/10/08.587 & 2009/11/15.725 \\
    M31N\,2009-11b & 63 & 2009/08/20.429 & 2009/11/15.725 \\
    M31N\,2010-06a & 13 & 2010/08/12.894 & 2010/08/21.998 \\
    M31N\,2010-06b & 13 & 2010/08/12.894 & 2010/08/21.998 \\
    M31N\,2010-06d & 13 & 2010/08/12.894 & 2010/08/21.998 \\
    M31N\,2010-07a & 13 & 2010/08/12.894 & 2010/08/21.998
    \enddata
  \end{deluxetable*}
\end{center}

\begin{center}
\begin{deluxetable*}{cccc}
 \tablecaption{X-ray and Spectroscopic observation epochs of M31 novae \label{XS}}
  \tablehead{
    \colhead{} & \multicolumn{2}{c}{Observation UT} & \colhead{} \\
    \colhead{Novae} & \colhead{Spectrum} & \colhead{X-Ray} & \colhead{Reference}
  }
  \startdata
  M31N\,2009-08e & 2009/09/18.014 & & \citealp{ATel2213} \\
  M31N\,2009-09a & 2009/09/18.18 & & \citealp{ATel2205} \\
  M31N\,2009-10a & 2009/10/09.91 & & \citealp{ATel2239} \\
  M31N\,2009-10b & 2009/10/09.820 & & \citealp{CBET1971p3} \\
                 & 2009/10/15.5   & & \citealp{CBET1980p1} \\
                 & 2009/10/19.936 & & \citealp{ATel2251} \\
  M31N\,2009-11a & 2009/11/25.32 & & \citealp{CBET2062} \\
  M31N\,2009-11b & 2009/11/08.35 & & \citealp{CBET2015p3} \\
                 & 2009/11/09.82 & & \citealp{ATel2290} \\
  M31N\,2009-11d & 2009/11/21.31 & & \citealp{CBET2058p3} \\
  M31N\,2009-11e & 2009/11/22.19 & & \citealp{CBET2061p5} \\
  M31N\,2010-01a & 2010/01/13.08 & & \citealp{CBET2127} \\
                 & 2010/01/15.09 & & \citealp{CBET2136} \\
  M31N\,2010-06a & 2010/07/19 &    & \citealp{CBET2391p3} \\
  M31N\,2010-06b & 2010/07/19 &    & \citealp{CBET2391p3} \\
  M31N\,2010-06d & 2010/08/08.86 & & \citealp{ATel2789} \\
  M31N\,2010-07a & 2010/07/19 &   & \citealp{CBET2391p3} \\
  M31N\,2010-09a & 2010/09/07.23 & & \citealp{ATel2834} \\
  M31N\,2010-09b & 2010/10/01.39 & & \citealp{ATel2898} \\
                 & 2010/10/06.40 & & \citealp{ATel2909} \\
  M31N\,2010-09c & 2010/09/14.23 & & \citealp{ATel2843} \\
  M31N\,2010-10a & 2010/10/06.38 & & \citealp{ATel2909} \\
  M31N\,2010-10b & 2010/10/28.31 & & \citealp{ATel3039} \\
                 & 2010/11/03.31 & & \citealp{ATel3039} \\
                 & 2010/11/11.28 & & \citealp{ATel3039} \\
  M31N\,2010-10c & 2010/10/17.35 & & \citealp{ATel2949} \\
  M31N\,2010-10d & 2010/10/30.31 & & \citealp{ATel2987} \\
  M31N\,2010-10e & 2010/11/04.32 & & \citealp{ATel3006} \\
                 & & 2010/11/15.16 & \citealp{ATel3038} \\
                 & & 2010/11/17.09 & \citealp{ATel3038} \\
  M31N\,2010-10f & 2010/11/13.285 & & \citealp{ATel3074} \\
                 & & 2010/11/03.04 & \citealp{ATel3013} \\
                 & & 2010/11/06.20  & \citealp{ATel3013} \\
                 & & 2010/11/07.06  & \citealp{ATel3013} \\
  M31N\,2010-11a & 2010/12/05.74 & & \citealp{ATel3068} \\
  M31N\,2010-12a & 2010/12/30.24 & & this paper \\
  M31N\,2010-12c & 2010/12/30.25 & & this paper
  \enddata
\end{deluxetable*}
\end{center}

\subsection{PTF}

PTF \citep{LKD2009} employs the 1.2-m Oschin Telescope at Palomar Observatory
with an array of twelve CCD chips, each having 4K$\times$2K pixels. One
of the chips is dead for unknown reasons. The pixel size
is $1\farcsec0$/pixel and thus the total FoV is 7.2 square degree. 

\subsubsection{Survey Observation}

The time coverage of PTF observations of M31 during the 2009 and 2010
seasons is shown in Figure \ref{Temp_Cover}.  The first observing season is
from 2009 August 18 to 2010 January 17 and the second is from 2010 June 30
to 2010 December 15.  From 2010 August 28 to 2010 September 7, PTF
observed M31 with a fast cadence of about ten minutes. All PTF images were
taken with a R-band filter.  The average seeing during the two seasons is
$2\farcsec5$ with a median value of $2\farcsec4$. We removed less than $1\%$ of the
images with seeing worse than $3\farcsec5$, because the high stellar density
in M31 leads to strong overlaps among adjacent objects under bad seeing
conditions.  Our final image database samples over seven hundred epochs on
7987 individual CCD images.

\subsubsection{Data Reduction and Photometry}

All images were processed through the PTF LBL pipeline \citep[see ][]{LKD2009},
obtaining astrometric solutions of $1\farcsec0$ accuracy. The calibration of
relative photometry proceeded as follows. We first use SExtractor to
make a source catalog for each image. We then choose the catalog with the largest number of
$14\,{\rm mag}<R<17\,{\rm mag}$ sources as reference. Then, other non-reference catalogs are
spatially matched to the reference using source positions and forced to
agree with the reference photometry after non-constant sources are removed
with a standard outlier procedure\footnote{{\it NIST/SMATCH e-Handbook of
Statistical Methods} at\\ http://www.itl.nist.gov/div898/handbook/}. The
resulting photometric uncertainty is usually around $0.05\,$mag. The absolute
calibration is achieved by matching the reference sources to sources in the
catalog presented in \citet{MOH2006}. 

Photometry is also done with SExtractor. Background is estimated in a
region of 20\farcsec$\times$20\farcsec\ centered at each target. Fluxes
are measured within a Kron-like elliptical aperture. The first order 
moment of each object determines the Kron radius $r$ \citep{Kron1980} and 
the second order moment determines ellipticity and position angle. The
aperture has an area of $6.25\pi r^2$. 
For a null detection, a 3-$\sigma$ limiting
magnitude is estimated. 

In order to assess the quality of photometric results from SExtractor,
we constructed light curves of three novae (M31N\,2009-09a,
M31N\,2009-10a and M31N\,2009-11a) with three methods: 1) the one we described
above; 2) aperture photometry with an aperture radius of seeing and
with constant local background; 3) aperture photometry
with an aperture radius of seeing and with local background estimation
by linear regression. All three methods show consistent magnitude
measurements. Hence we adopt the first method. 

\subsubsection{Photometric and Spectroscopic Follow-up}

Photometric follow-up of M31N\,2010-11a was triggered for the Palomar
60-inch telescope \citep{CFM2006} on 2010 December 4. On 2010 December 30, we also
used the Low Resolution Imaging Spectrometer \citep[LRIS; ][]{OCC1995} mounted on the Keck I
telescope to obtain spectra of M31N\,2010-12a and 2010-12c. These
spectra were obtained with a configuration of the 560 dichroic, the
400/3400 grating in the blue side and the 400/8500 grating in the red
side with a central wavelength of $7800\,\AA$. 

\subsection{{\it GALEX}}
\label{galex}

{\it GALEX} is a wide-field imaging UV space telescope that was originally
launched with near-UV (NUV and far-UV (FUV) detectors. Unfortunately, the 
FUV detector failed before the start of the PTF survey of M31
The NV detector effective wavelength is $2316\,$\AA\, and the band
width is $1060\,$\AA\,\citep{MFS2005,MCB2007}. This channel has a FoV of $1.2$
square degree \citep{MCB2007}. 

\subsubsection{{\it GALEX} Observation Summary}

The visibility of M31 for {\it GALEX} is defined by the Sun, Moon and Earth-limb
 constraints that limit the observations more than for ground-based
observatories.  The result is that {\it GALEX} was able to intensively
monitor M31 in several campaigns that overlap with the PTF campaigns of 2009 and
2010 (2009 August 08 to 2009 September 04; 2009 October 02 to 2009 October
21; 2009 November 15 to 2009 November 22; 2010 August 12 to 2010 August
21). Figure \ref{Temp_Cover} illustrates {\it GALEX} temporal
coverage and its overlap with PTF.  A field centered in M31 is monitored
daily, while ten outskirt fields in M31 are observed every other
day. So we have fifty-one epochs of the center field while about 22$-$27 epochs
of the outskirt fields. The exposures were all taken
during single orbit pointings and thus have durations close to 1500 seconds
with a range from 740 seconds to 1703 seconds.  Table \ref{obs_sum}
summarizes parameters of observations.

\subsubsection{Photometry}

Photometry was done with an aperture
with a radius of six arcseconds at each nova position, and was
calibrated using the standard {\it GALEX} zero-points to the AB magnitude
system \citep{MCB2007}. Eight novae were detected; see Table \ref{galexobs}.

\subsection{Data from Other Sources}
For completeness in optical and UV bands, we incorporate photometric measurements of novae from ATel and CBET reports (see the reference column
in Table \ref{Nova_List}). This includes optical photometry in R, B, V, r', i' and H$\alpha$ bands as well as unfiltered data. 
We also include data published in \citet{SBD2011} where several novae in 2009 are well-sampled. 
The UV data in ATel reports are mainly obtained by {\it Swift} \citep{GCG2004}. For consistency of data between {\it GALEX} and {\it Swift}, 
all {\it Swift} data are calibrated onto the AB magnitude system \citep{SHR2010}. 
When data from different sources are used, due to difference of filters, data may have
systematical errors among different projects. 

In addition, we include X-ray and spectroscopic information of novae in M31 from {\it Swift} from either ATel reports or our own analysis (see Table \ref{XS}). 

\subsection{Missed Novae}
During the two seasons presented here, we did not find three announced
novae in the PTF archive: M31N\,2009-10c, M31N\,2009-11c and M31N\,2010-07b. 

M31N\,2009-10c took place at $\alpha=00^h42^m45^s8$, 
$\delta=+41^\circ15'57\farcsec$ (J2000), less than thirty arcseconds from the center of M31. This nova reached maximum light, 
i.e., $R=17.2$, on Oct 9, 2009 \citep{ATel2234}. PTF observed M31 on Oct 10, 2009. In the vicinity of the nova position, if we fit a constant background, the
mean count of $\sim 50000$ with a standard deviation of $\sigma\sim6000$. Given
 that the seeing then was about three arcseconds and that the zeropoint
is $27.6$, the 3-$\sigma$ limiting magnitude is $15.2$. If the background is fit with a plane, the standard deviation is roughly $1500$ and the limiting magnitude
is $16.8$. Thus we did not detect this nova in the images. In the
subtraction image, the circle of thirty arcseconds in radius in the center of M31
is excluded in the subtraction algorithm where 
the algorithm does not work well. As a result, we missed this nova in our PTF data. 

There are gaps of forty arcseconds between adjacent CCDs. 
M31N\,2009-11c unfortunately fell in the gap between two adjacent
CCDs.

M31N\,2010-07b had a peak luminosity of $R=20.7\,$mag on
2010/06/04.135 and slowly decayed \citep{CBET2411}, which is 
below the single-image detection threshold of PTF. The second
observing season of PTF started at the end of June (see Figure
\ref{obs_sum}), so we do not have enough images to co-add for
detection of this nova.  

Admittedly, we did not carry out a real-time search of novae in M31
in the first two seasons of PTF. Our nova sample inherits the
sensitivity and incompleteness of those projects that discovered these
novae. 

\subsection{Special Novae}
\subsubsection{M31\,2010-09c}
This nova has a faint and close neighbor with R$=19.85\pm0.2$ within two
arcseconds.  In SDSS Data Release 8, we found two nearby sources within two
arcseconds.  After color transformation \citep{JGA2006}, the neighbors have
R$=20.53$ and R$=20.33$, giving a total magnitude of $19.67$ that is
consistent with our measurements. The uncertainty caused by the neighbors
has been included in the nova light curve.  The spectroscopic confirmation
in \citep{ATel2843} identified the M31N\,2010-09c as a real nova of Fe~II
type.

\subsubsection{M31\,2010-10f}
This nova is spatially associated with the M31 globular cluster Bol~126
\citep{WSB1985}.  In Figure~\ref{Bol126}, we clearly see a brightening of
the globular cluster by about one magnitude on around 2010 October 10
(Julian Date: 2455480). The luminosity of the globular cluster before the
nova is $16.8\pm0.2$. The light curve of the nova is obtained by
subtracting the globular cluster brightness.  One possible reason for the
large uncertainty in the photometry measurements is that the nova is only
a few pixels away from the edge of the CCD. 

\subsubsection{M31N\,2010-01a and M31N\,2010-12c}
These two novae occurred within one arcsecond of each other, a separation
which is less than  the typical spatial resolution of PTF ($\sim$1\farcsec0). 
A highly accurate astrometric solution shows that they are
indeed two separate objects \citep{CBET2610p1}. 

\begin{figure*}
\begin{center}
\includegraphics[width=0.8\textwidth]{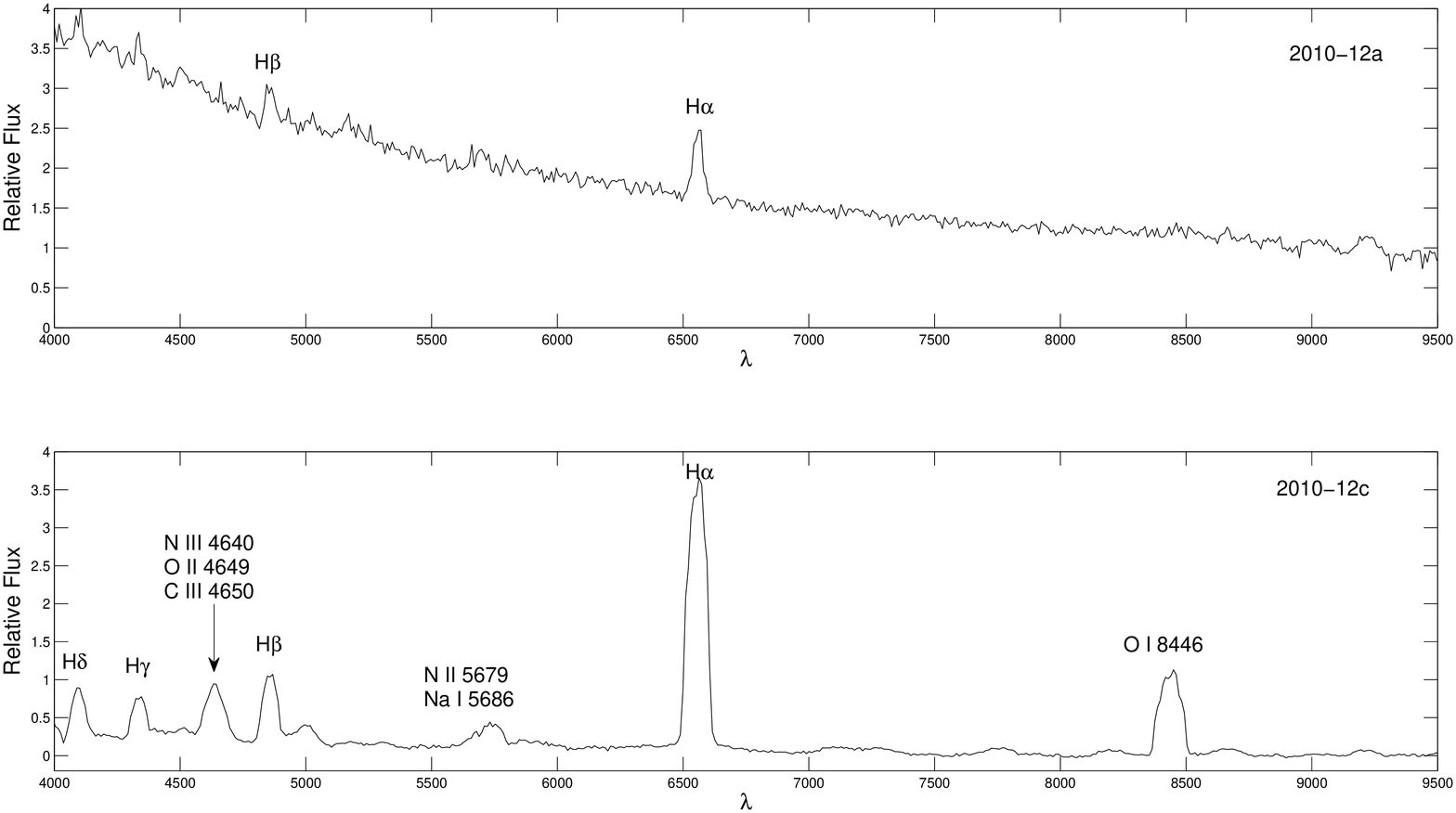}
\caption{Spectra of M31N\,2010-12a (top) and M31N\,2010-12c (bottom),
obtained with LRIS on Keck-I telescope on December 30th, 2010.
\label{spec}}
\end{center}
\end{figure*}

We obtained a spectrum of 2010-12c with LRIS on 2010 December 30 (Figure
\ref{spec}, bottom panel).  The spectrum shows a weak continuum superposed
with many emission lines, illustrating that the nova has already entered
the nebular phase.  We identified H$\alpha$, H$\beta$, H$\gamma$,
H$\delta$, OI 8446 and several He and N lines.  H$\delta$ has a slight P
Cygni profile.  The emission lines suggest the nova is of the He/N type.
After fitting the emission lines with a Gaussian profile, we find that
H$\alpha$ is centered at $6556$~\AA, H$\beta$ at $4858$~\AA, H$\delta$ at
$4338$~\AA, and OI 8446 at $8437$~\AA, suggesting a blueshifted velocity in
the range from $-200$ to $-300$ km s$^{-1}$. On the other hand, the
systemic velocity of M31 is $-300$ km s$^{-1}$. The nova lies about $2'$
northeast of the center of M31. The HI 21~cm line \citep{CCF2009,CLW2010}
suggests a rotation velocity of $100$ km s$^{-1}$ receding from us.
However, the nova is apparently close to the center of the galaxy, and therefore may
not share the motion of the disk measured by the HI 21cm line.  Another
measurement of a nearby planetary nebula \citep{MMD2006} at
$\alpha=00^h42^m57^s.4$ and $\delta=+41^{\circ}17'26\farcsec$ (J2000) suggests a radial
velocity of $-300$ km s$^{-1}$ along the line of sight.  Consequently, we
conclude that 2010-12c is a real nova in M31.

Additionally, these two novae have two more historical neighboring novae within $10\farcsec$: nova 30
(discovered in 1986) from \citet{CFN1987} and
M31N\,2009-08a. Generally speaking, we observed more novae in the central
region of M31 than in outskirts. This might result
from the high stellar density in the center of the galaxy. 


\section{Light Curves of Classical Novae}
\label{section_LC}
\begin{figure*}
\begin{center}
\includegraphics[width=\textwidth]{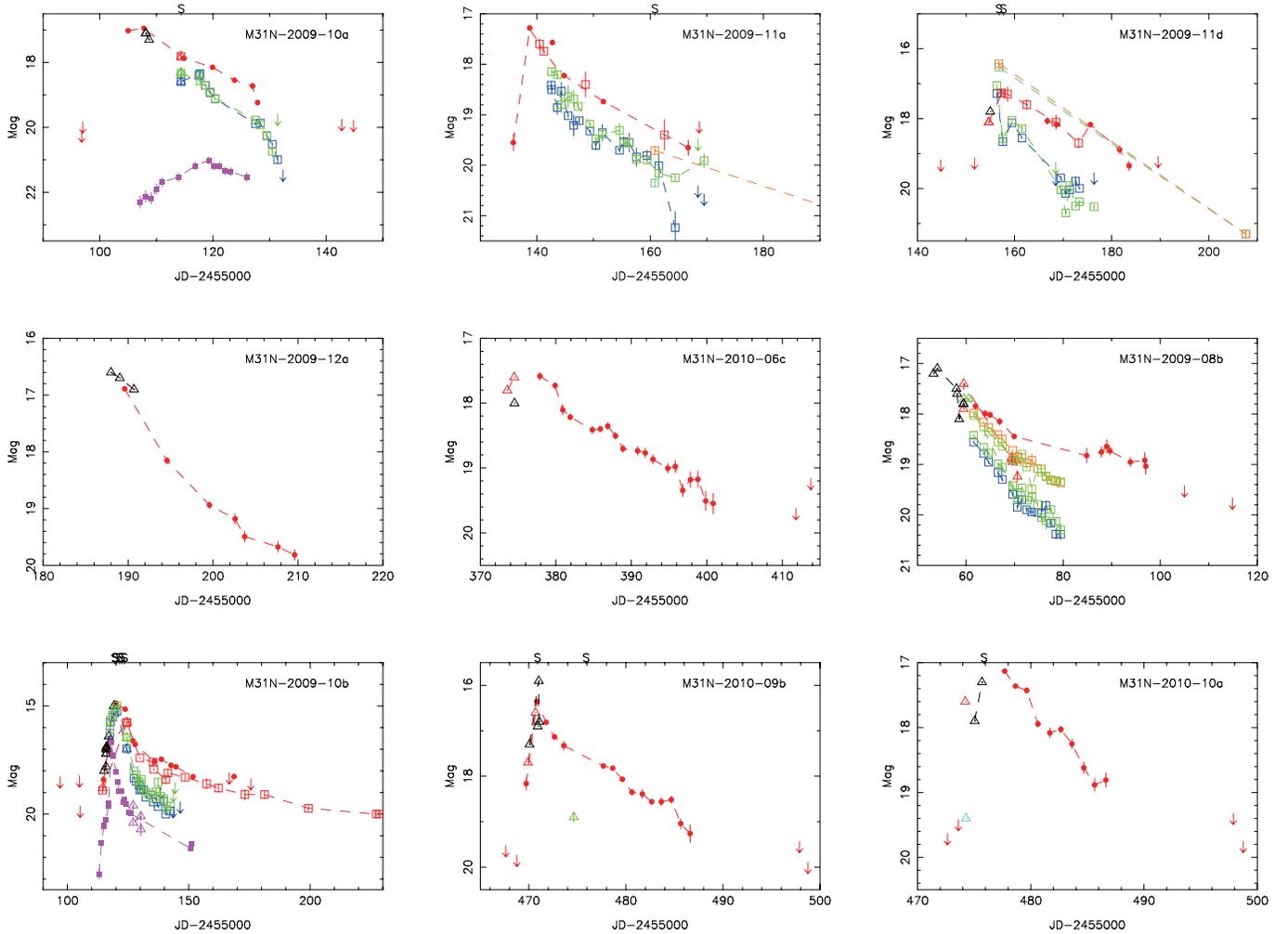}
\caption{Light curves of nine smoothly declining novae.  
Different symbol shapes mean data from different sources. 
PTF and {\it GALEX} data are shown in filled circles and rectangles, respectively.
Data collected from ATel and CBET reports are in empty triangles and data 
read from \citet{SDH2011} are in empty rectangles. 
Data in different filters are illustrated by colors. R band data are in red; B band data
in blue; V band data in green; r$'$ band data in orange; i$'$ band data in Chartreuse; 
H$\alpha$ narrow band data in cyan; NUV band data in magenta; and unfiltered data in black. 
In the top axis, an ``X'' indicates an X-ray observation at that time and an ``S''
indicates a spectrum of the nova taken at that time.
\label{lc2}}
\end{center}
\end{figure*}

\begin{figure*}
\begin{center}
\includegraphics[width=0.8\textwidth]{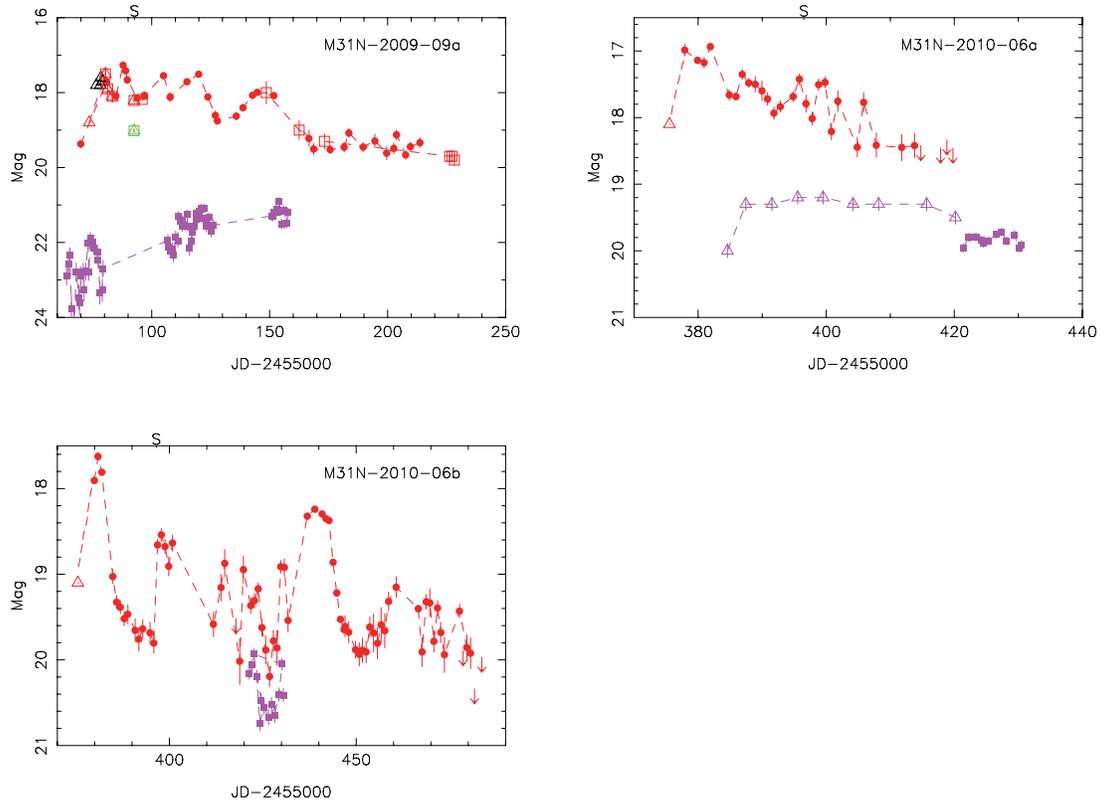}
\caption{Light curves of novae with jittering decay.  
Symbol designation is the same as that in Figure \ref{lc2}. 
\label{lc4}}
\end{center}
\end{figure*}

\begin{figure*}
\begin{center}
\includegraphics[angle=270,width=0.8\textwidth]{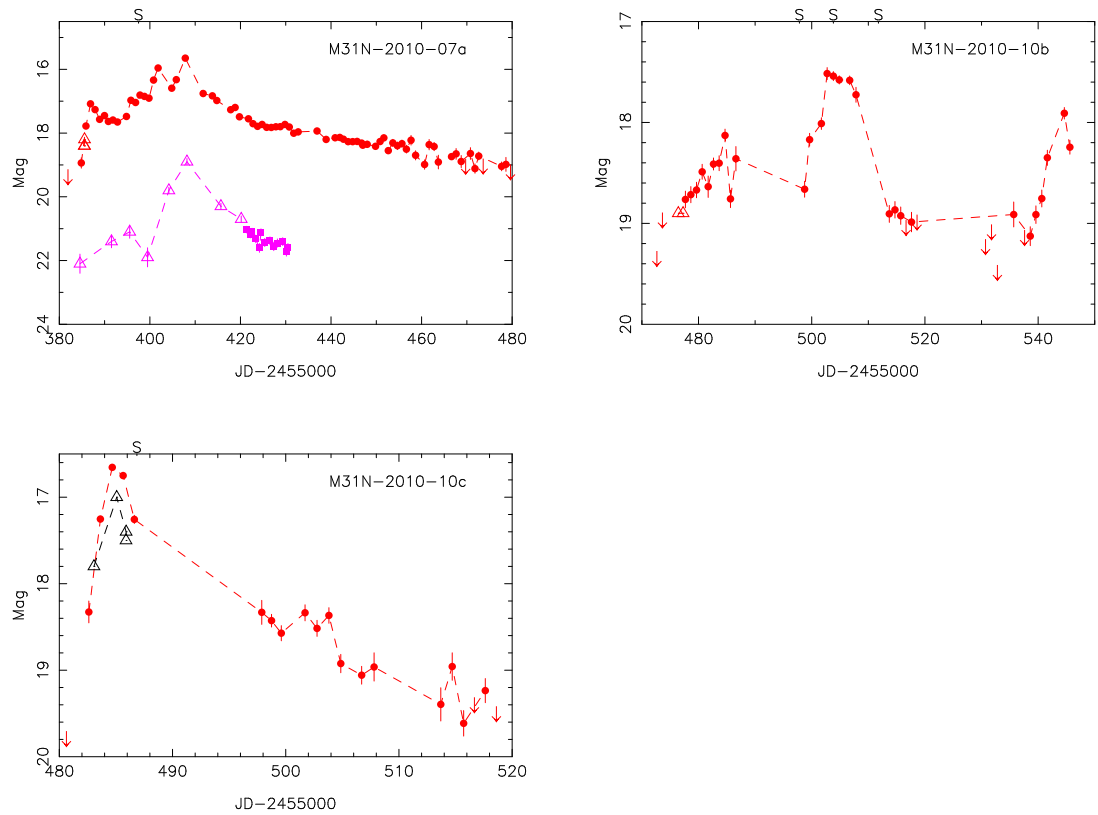}
\caption{Light curves whose rise stages are well-sampled.  
Symbol designation is the same as that in Figure \ref{lc2}. 
\label{lc5}}
\end{center}
\end{figure*}

\begin{figure*}
\begin{center}
\includegraphics[angle=270,width=0.8\textwidth]{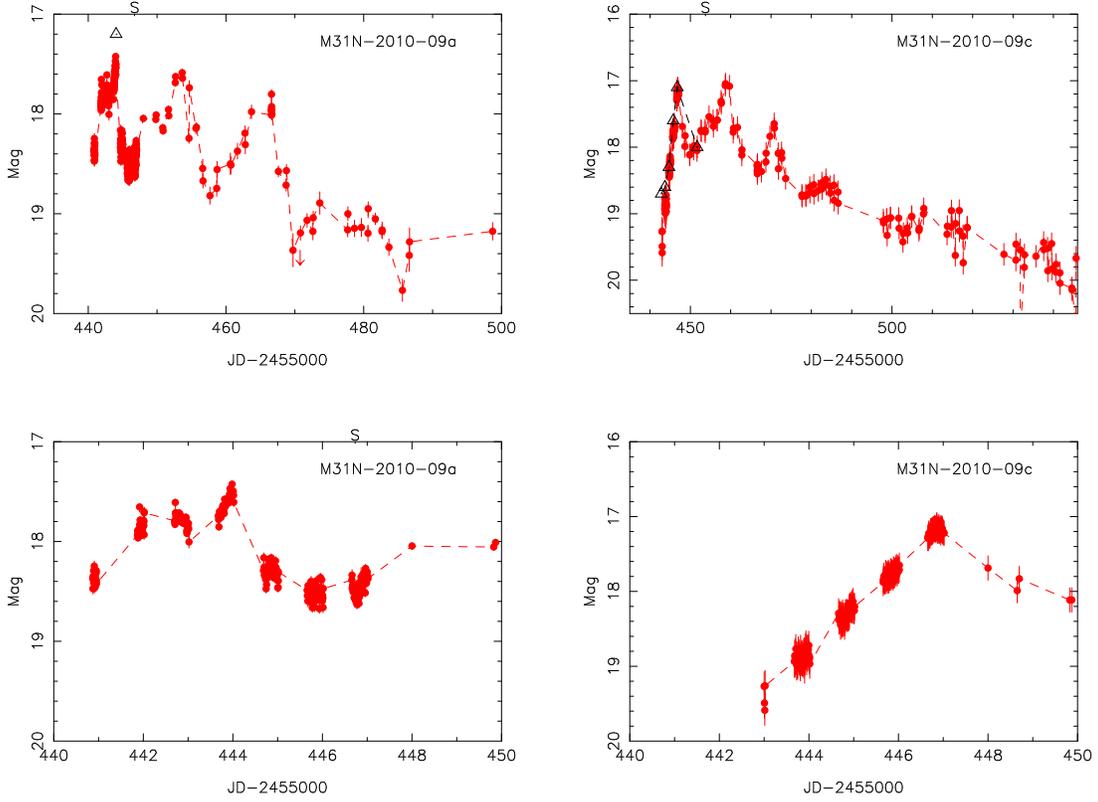}
\caption{M31N\,2010-09a and M31N\,2010-09c were observed during the PTF
10min-cadence experiments.  In the top two panels, we show both full light
curves. In the bottom two panels, rising phases are highlighted. 
Symbol designation is the same as that in Figure \ref{lc2}. 
\label{rise}}
\end{center}
\end{figure*}

\begin{figure*}
\begin{center}
\includegraphics[width=0.8\textwidth]{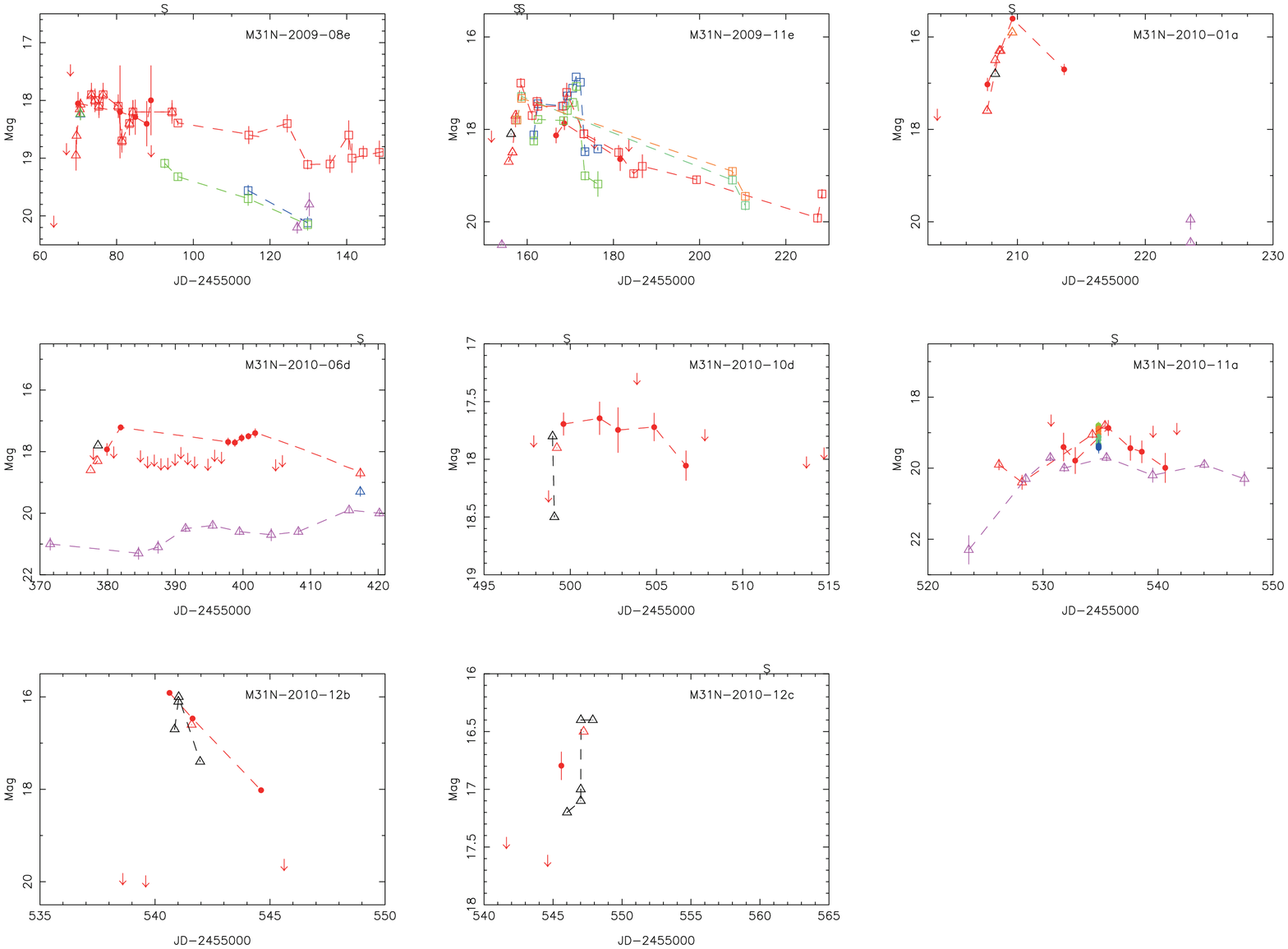}
\caption{The light curves of under-sampled novae.  
Symbol designation is the same as that in Figure \ref{lc2}. 
\label{lc6}}
\end{center}
\end{figure*}

Except three recurrent events, light curves of twenty-six classical novae are presented in Figure~\ref{lc2} to \ref{lc6}. 
An electronic version of the PTF and {\it GALEX} data can be found online\footnote{http://www.astro.caltech.edu/$\sim$ycao/m31novae/index.html}
in a machine-readable format.  An example of M31N\,2010-12a is illustrated
in Table~\ref{2010-12a}. 

\begin{figure}
\begin{center}
\includegraphics[angle=270,width=0.5\textwidth]{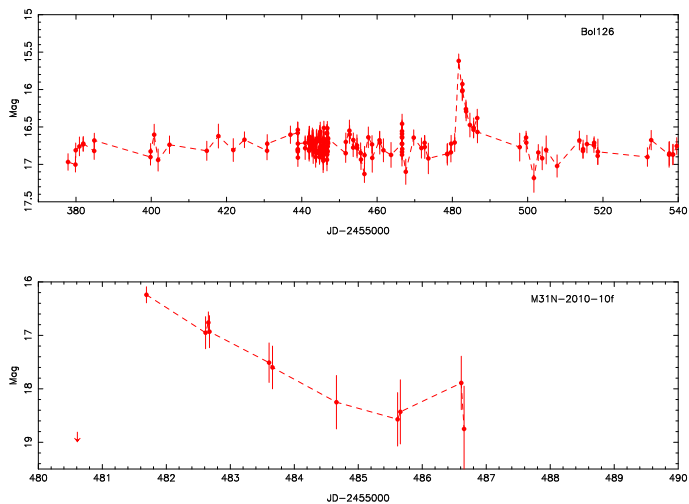}
\caption{The light curve of the globular cluster Bol~126 is shown in the
top panel. The subtracted light curve of M31N\,2010-10f is presented in the
bottom panel.  
\label{Bol126}}
\end{center}
\end{figure}

The nova light curves are presented in groups based on their features
described below. Figures~\ref{lc2} presents novae with smooth decline. 
Figure~\ref{lc4} illustrates novae with jittery decline.  Novae with
well-sampled rise stages appear in Figure~\ref{lc5}. Figure~\ref{rise}
highlights two novae recorded during the PTF 10min-cadence observations
where their rise histories were recorded in great detail.  Figure~\ref{Bol126} 
shows the light curves of the only globular cluster nova
M31N\,2010-10f and its host Bol~126.  All other under-sampled light curves
are collected in Figure~\ref{lc6}. 

Table~\ref{NovaProperties} summarizes the photometric and morphological features of the novae. 

%

\begin{deluxetable}{ccc}
    \tablecaption{Photometric measurements of M31N\,2010-12a\label{2010-12a}}
    \tablehead{\colhead{Julian Date} & \colhead{Magnitude} & \colhead{Magnitude Error\tablenotemark{1}}}
    \startdata
     2455530.702 & 20.190 & 99.000 \\
     2455531.770 & 16.763 &  0.030 \\
     2455532.780 & 15.581 &  0.023 \\
     2455532.825 & 15.561 &  0.025 \\
     2455535.681 & 16.013 &  0.024 \\
     2455537.578 & 16.325 &  0.024 \\
     2455537.622 & 16.346 &  0.025 \\
     2455538.577 & 16.526 &  0.024 \\
     2455538.621 & 16.496 &  0.025 \\
     2455539.577 & 16.636 &  0.026 \\
     2455539.620 & 16.638 &  0.025 \\
     2455540.615 & 16.721 &  0.027 \\
     2455540.659 & 16.776 &  0.026 \\
     2455541.613 & 16.817 &  0.027 \\
     2455541.656 & 16.899 &  0.031 \\
     2455544.593 & 17.243 &  0.029 \\
     2455544.639 & 17.175 &  0.032 \\
     2455545.593 & 17.283 &  0.038 \\
     2455545.637 & 17.303 &  0.034
    \enddata
    \tablenotetext{1}{99.000 in Magnitude Error indicates that a 3$\sigma$ upper limit is given in Magnitude.}
\end{deluxetable}

\subsection{Morphological classification}
The shape of each light curve can be characterized by several parameters:
the rise rate, the rise time to the maximum magnitude, 
the rise behavior, the maximum magnitude, the decline rate that is
represented by $t_n (n=1,\ 2,\ 3,\ \dots)$ defined as the time interval
during which a nova decays from its maximum by $n$ magnitudes, and 
the decline behavior.  Since PTF samples the M31 nova luminosity
distribution down to M$_R \sim-4.5$ (see \S1) and the peak absolute
magnitude of a nova can range from $-5$ to $-9$, we use $t_2$ to indicate the
decline rate. In our measurements, $t_2$ is estimated by linear
interpolation between adjacent measurements.  Other parameters are defined
as follows: the rise rate is the mean rate from the first detection to the
first local maximum; the rise time is the interval
from first detection in PTF to its global maximum. All measurements are done with
PTF R-band data.

For novae inside our Galaxy, \citet{SSH2010} classified the diverse light
curves morphologically according to their evolution of decline. With a median
coverage extending to 8.0 mag below peak, the authors grouped nova light
curves into seven classes. In their sample of 93 nova light curves, $38\%$
decline smoothly, $21\%$ have plateau phase, $18\%$ show dust dips,
$1\%$ rebrighten with a cusp-shaped secondary maxima, $4\%$ oscillate
quasi-sinusoidally on smooth decline, $2\%$ are flat-topped
and $16\%$ have irregular jitters or flared during decline. 
However, PTF is unable to follow M31 novae to such late times in the light
curve evolution.  We, instead, roughly classify our M31 novae into two
major classes: S-class for smooth light curves and J-class for jittery
light curves or for light curves with flares superposed on the decline.
Statistically in our nova sample, $60\%$ show S-class decline and $40\%$
belong to the J-class. Given that our sample is quite small, this result is
roughly consistent with the Galactic population. We do not see any flat-topped
nova. 

Similarly, due to the high cadences, PTF sampled the rising stages of several novae very well.
This enables us to employ the same S-J classification to describe the
rising parts of the light
curves (see Table \ref{NovaProperties}). We find that the S-class novae rise very fast (e.g., M31N\,2010-09b rose within two days)
while the J-class novae rise more slowly (e.g., M31N\,2010-10b took thirty days to reach its maximum). 
But we do not see any relation between the rise time and peak magnitudes. Luminous M31N\,2010-09b
and subluminous M31N\,2009-11a rose rapidly while luminous M31N\,2010-07a and subluminous
M31N\,2009-09a both rose very slowly.

During a very-high-cadence (tens of minutes) experiment in PTF, we
observed the rising stages of M31N\,2010-09a and M31N\,2010-09c in
unprecedented detail, as shown in the bottom panels of Figure \ref{rise}. 
The rise of both
novae is smooth over a timescale of tens of minutes. 

\subsection{Super-Eddington Phase}
Some novae around maximum exhibit a super-Eddington phase.  Theorists
\citep[e.g.,][]{Shaviv2001} model this phase with a porous structure in the
nova wind that reduces the effective opacity and raises the Eddington
luminosity. \citet{KH2005} produced a light-curve model based on this idea
that successfully matched the super-Eddington phase of Nova V1974 Cyg, and
demonstrated the importance of the relationship between the optical and UV
light curves.  They isolated a portion of the UV continuum near 1455~\AA\,
that had a duration that was well correlated with the decline rate of the
nova in the optical.  This continuum UV light curve combined with the model
of how the UV light is reprocessed into the optical light curve allows an
estimate of the precursor WD mass and distance to be made, based on certain
assumptions about the chemical composition of the envelope \citep{KH2007}.

One might assume that the clumpy wind responsible for the lower opacity and
higher Eddington luminosities of the brightest novae might produce other
observable signatures.  Small-scale structure in the rising light curve
could be observed as the clumps are non-uniformly heated and accelerated by
the ongoing eruption and as radiation is diffused through them.  We have
very high-cadence observations of the rising phases of two novae (2010-09a
and 2010-09c), both of which have a peak brightness fainter than -7.5 and thus
are probably not super-Eddington novae.  Nonetheless, while 2010-09c
exhibits a very smooth and steady rise and a smooth decline, consistent
with a uniform nova wind, 2010-09a has considerable structure in its light
curve with multiple peaks both near and after the maximum light.  Our
observations of 2009-09a, 2010-07a, 2009-09a and 2009-09c also show several
peaks around maxima.  The multiple peaks could arise when different winds
are blown out from the center successively during an outburst.  Each peak
is formed when the pseudo-photosphere meets the shock wave at the head of
the corresponding wind.  We conclude that structures in the nova wind and
thus in the light curve can arise regardless of the luminosity of the
outburst.

\section{NOVAE IN GLOBULAR CLUSTERS}
\label{section_gc}
M31N\,2010-10f was spatially associated with the globular cluster (GC) Bol
126 to within one arcsecond. A supersoft X-ray source (SSS) was also
reported in the vicinity by \citet{ATel3019}.  This event is similar to
M31N\,2007-06b \citep{SQ2007} in Bol~111 which was confirmed
spectroscopically during the outburst. A SSS counterpart was also found for
this object \citep{HPH2009}.  They are the only two events in M31 that are
associated with GCs. 

Several previous searches for GC novae obtained null results
\citep{CTP1990,TCS1992}.  We examined our PTF image database at the
positions of all confirmed M31 GCs from the Revised Bologna Catalog
\citep[V.4, Dec 2009,][]{GFB2004}. Magnitudes of these GCs range from 
R$=$13\,mag to R$=$20\,mag. We do not find any other GC novae
candidates.  Given an effective survey time of M31 in PTF approximating one
year, we obtain a GC nova rate of about 1~yr$^{-1}$.

Is the nova rate enhanced in GCs? The total stellar mass of M31 is $\sim
6\times10^{10}\sm$ \citep{TTT2007}.  Using the highest published nova rate
for M31 of 65~yr$^{-1}$ from \citet{DBK2006}, we estimate a maximum global
specific nova rate of $\sim 1 /(10^{9}\sm)$~yr$^{-1}$ in M31.  Given a
typical GC mass of $10^6\sm$ and about seven hundred confirmed M31 GCs, the nova rate
in the M31 GC system is $\sim 1/(10^9\sm)$~yr$^{-1}$. We conclude that the
M31 GC specific nova rate is not significantly higher than the M31 galaxy
specific nova rate.

LMXBs -- mass transferring binaries with a neutron star receiving matter from
a companion -- can be viewed, in some ways, as surrogates for CNe and RNe. 
Since the early days of X-ray astronomy it has been known that at least in 
our Galaxy that the brightest LMXBs (per unit mass) are nearly two orders of
magnitude larger than that for the disk of the Milky Way. This trend appears 
to be even more acute for M31 \citep{TP2004}. In part this may be because M31
possesses a more extensive cluster system than our own Galaxy. Thus, naively,
we would expect far more CNe and RNe in the globular cluster system of M31. 
Over the period of investigation (2009-2010), a total of three RNe were found 
in the disk/bulge of M31. If we assume the mass fraction of globular clusters 
is $\sim 1\%$, we would have expected to see a similar number in the
cluster system. Within small number systems, the detection of RN associated
with the cluster would be consistent with this expectation. Separately, given
that seven hundred CNe are known, over 2009 and 2010 we should have seen 
many tens of CNe in the globular clusters.
but only one was found. Indeed, over the last hundred years only one CN 
that has been associated with a globular cluster of M31 has been 
spectroscopically
confirmed \citep{SQ2007}. In summary we cannot arrive at a sensible conclusion
with the presently available data. 

\section{Recurrent Novae}
\label{section_RN}
\begin{figure*}
\begin{center}
\includegraphics[angle=270,width=0.8\textwidth]{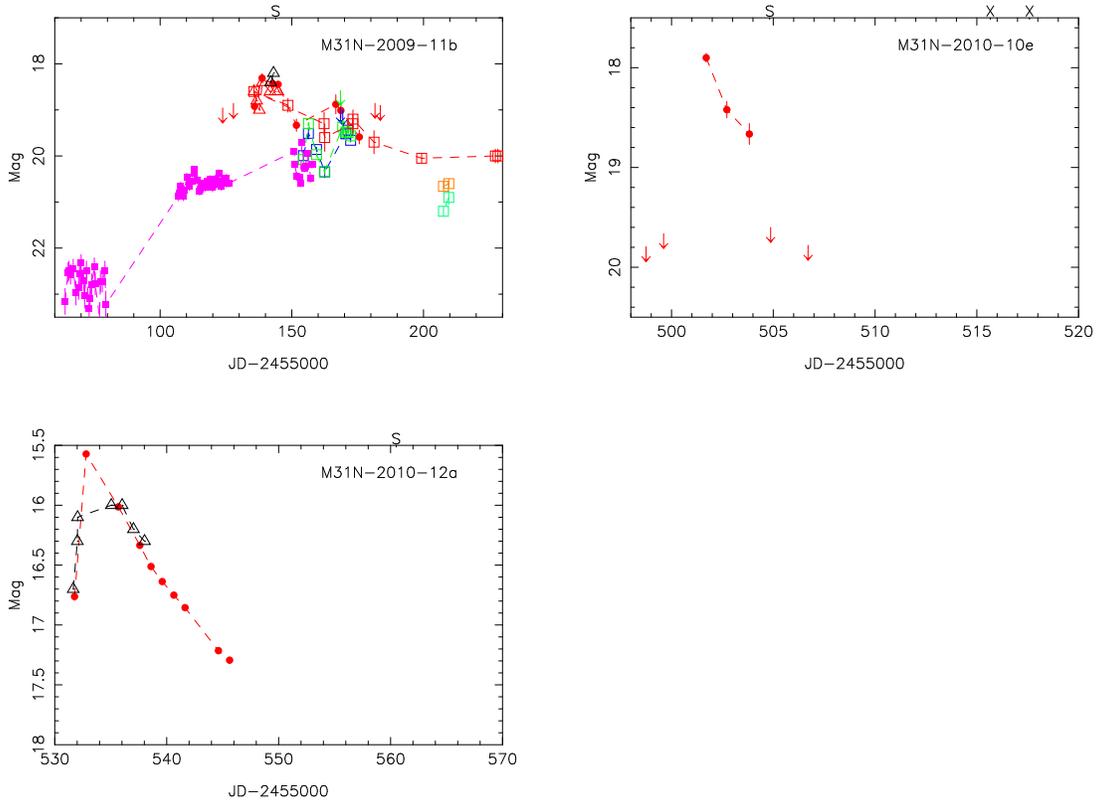}
\caption{Light curves of 3 RNe detected in PTF.
Symbol designation is the same as that in Figure \ref{lc2}. 
\label{lc1}}
\end{center}
\end{figure*}

\begin{figure}
\begin{center}
\includegraphics[width=0.5\textwidth]{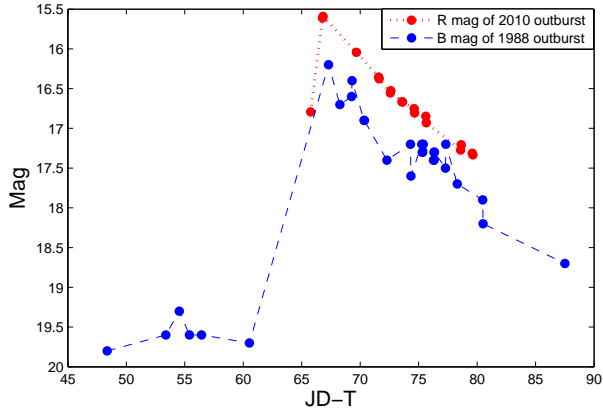}
\caption{A comparison of the light curves of outbursts of PT And in 1988
and 2010 is shown in the figure, where $T=2455466$ for the 2010
outburst and $T=2455100$ for the 1988 one. Though with different
filters, the light curve of the recent
outburst (M31N\,2010-12a) resembles that of the 1988 outburst.
\label{PT_And}}
\end{center}
\end{figure}

\begin{deluxetable*}{ccccccc}
    \tablecaption{Properties of PTF well-sampled nova light curves\label{NovaProperties}}
    \tablehead{
        \colhead{}     & \colhead{Rising rate} & \colhead{Rising Time} & \colhead{Rising} & \colhead{$t_2$} & \colhead{Decline} & \colhead{R-Band Peak} \\
        \colhead{Nova} & \colhead{(mag/day)} & \colhead{(days)} & \colhead{Morphology} & \colhead{(days)} &
        \colhead{Morphology} & \colhead{Magnitude}
    }
    \startdata
    2009-08b & ... & ... & ... & $>$35 & J & 17.76 \\
    2009-09a & ... & 26  &  J  &  79   & J & 17.26 \\
    2009-10a & ... & ... & ... &  20   & S & 16.95 \\
    2009-10b & 0.6 &  5  &  S  &  8    & S & 14.84 \\
    2009-11a & 0.8 &  3  &  S  &  22   & S & 17.23 \\
    2009-11b & ... & ... & ... & $>$37 & J & 18.25 \\
    2009-11d & ... & ... & ... & $>$17 & J & 17.72\tablenotemark{1}\\
    2009-12a & ... & ... & ... &  10   & S & 16.88 \\
    2010-06a & ... & ... & ... & $>$31 & J & 16.88 \\
    2010-06b & 0.4 & ... & ... &  8    & J & 17.62 \\
    2010-06c & ... & ... & ... &  19   & S & 17.57 \\
    2010-07a & 0.7 & 26  &  J  &  15   & S & 15.65 \\
    2010-09a & 0.5 & 14  &  J  &  40   & J & 17.42 \\
    2010-09b & 1.6 &  2  &  S  &  10   & S & 16.36 \\
    2010-09c & 0.4 & 18  &  J  &  57   & J & 17.04 \\
    2010-10a & ... & ... & ... & $>$9  & S & 17.03 \\
    2010-10b & ... & 30  & ... & $>$41 & ... & 17.44 \\
    2010-10c & 1.0 &  3  &  S  &  20   & S & 16.65 \\
    2010-10e & ... & ... & ... & $>$5  & S & 17.84 \\
    2010-12a & 1.0 &  3  &  S  & $>$13 & S & 15.56 \\
    2010-12b & ... & ... & ... &  3    & S & 15.70 \\
    \enddata
    \tablenotetext{1}{The maximum magnitude is from ATel reports because PTF missed the peak magnitude.}
\end{deluxetable*}

According to the catalog maintained by \citet{Pietsch2010}, 26 RNe in M31 are known. During the
two seasons we report on, three recurrent events were reported:
M31N\,2009-11b, M31N\,2010-10e and M31N\,2010-12a. Figure~\ref{lc1}
presents their light curves.  Besides these, we
examined the PTF archive at the positions of more than 800 old novae in M31
from the list of \citet{Pietsch2010} in order to find more recurrent
events.  None were found.  Therefore, given an effective observing time
of roughly one year in PTF, the recurrent nova rate in M31 is 
$\sim3\,$novae per year.

We examined M31N\,2010-12a in some detail, as another outburst of PT And.
This object has recorded outbursts in 1957, 1983, 1986, 1988 and 1998.
\citet{AZ2000} collected its previous outbursts and found that all light
curves can be fit by one template. The average decline rate of previous
outbursts is 0.09 to 0.11 mag per day. The light curve of this most recent
outburst (M31N\,2010-12a) can also be fit by the template. Figure
\ref{PT_And} shows the similarity between light curves of the recent and
1988 outbursts. The decline rate of the recent outburst is around
0.10 mag per day consistent with previous ones. This invariance property is
also observed in Galactic RNe \citep{Schaefer2010}. Such characteristics of
RNe suggests that RN outbursts only depend on system parameters like the WD
mass and its composition. 

Because of its luminous and frequent outbursts, PT And was assumed to be a
dwarf nova inside the Milky Way \citep{AZ2000}. However, no direct evidence of this object's
nature was available until we obtained a spectrum of its most recent
outburst with LRIS in Keck-I telescope. This spectrum (see the top panel of
Figure~\ref{spec}) shows a very blue continuum superposed with prominent
H$\alpha$ and H$\beta$ lines as well as several weak Fe lines. Fitting the
red part of the spectrum with a Planck function yields a temperature of
$7\sim8\times10^{3}\,K$.  We fit the H$\alpha$ and H$\beta$ lines with Gaussian profiles and
obtained the following observed wavelengths: H$\alpha$ is centered at
$\lambda=6558\,$\AA and H$\beta$ at $\lambda=4857\,$\AA.  These correspond to
blue-shifts with radial velocities of $\sim -200\,$km s$^{-1}$ and $\sim
-300\,$km s$^{-1}$.  The systemic velocity of M31 is $\sim-300\,$km s$^{-1}$.
We assume that this object is not located in the disk of M31, thus we do
not take the rotation velocity of M31 disk into account. We conclude that
the blue-shifts are roughly consistent with the motion of M31. Our spectrum
evidently supports the conclusion that PT And resides in M31 instead of in
the Milky Way.  This can also be tested by observing its proper motion. It is
of only galactic sources that we may observe proper motion. 
In addition, the existence of weak Fe lines suggests the
nova is of Fe~II type.

The association of PT And with M31 leads to a peak absolute magnitude
of about -9, which makes it the brightest RN ever known. 

The decline rates of RNe vary over a large range in M31. M31N\,2009-11b has
$t_2>37\,$days and M31N\,2010-10e has $t_2>5\,$days.  Consider that 2010-12a
has a decline rate of $\sim 0.1\,$mag/day, so its $t_2$ is possibly around 
twenty days. The $t_2$ values of Galactic RNe also range from one day to 
fifty days \citep{Schaefer2010}.

\section{UV Light Curves of Novae}
\label{section_UV}

Theoretically, the UV emission after the optical maximum light is characterized
by a delayed UV peak after the optical peak, because most UV photons
are absorbed and re-emitted in the optical in the outer envelope
\citep{KH1994, GK1998, CAG2002}. 
However in our sample, we see some deviation from this general picture:
2009-10b has a peak in the {\it GALEX} NUV over two days prior to its peak
in the optical (see Figure \ref{2009-10b}).  The recurrent nova 2009-11b
was seen in the {\it GALEX} NUV about forty days before it was visible in
optical bands (although this could be due to cadence and weather, see
Figure~\ref{lc1}). The nova 2010-06d was detected in {\it Swift}
\citep{ATel2713} about six days before optical detection (see Figure
\ref{lc6}). {\it Swift} also detected 2010-11a several days before optical
detection \citep[][; also Figure \ref{lc6}]{ATel3061,ATel3066}. For 2010-07a,
the optical and UV maxima were nearly simultaneous as measured by PTF and
by {\it Swift} \citep[][see Figure~\ref{lc5}]{ATel2727}.

Admittedly the detection in the optical and UV involves many factors like
survey sensitivities, temporal coverage and observing weather.   In a few
cases, however, it is clear that the UV peak coincides with or precedes the
optical peak.  These cases demonstrate that not all UV photons are
re-processed and that the absorption in the UV will vary considerably
depending on factors such as physical geometry or chemical abundances in
the envelope.  These variations will produce a range of optical/UV peak
offsets and luminosity ratios.  We also point out that multi-band optical
data show that three of these UV novae have red colors: 2009-10b has
$R-V\sim 1.4$ at decline \citep{SDH2011}; 2009-11b has $R-V\sim 0.7$ at
decline \citep{SDH2011}; 2010-11a has $g-i\sim0.2$ around its optical peak
from our photometric follow-up.

\subsection{M31N\,2009-10b and M31N\,2010-11a}\label{UV2009-10b}

The novae 2009-10b and 2010-11a are two peculiar cases in that their UV
peaks lead the optical maxima by a few days.
 Moreover, at the early decay stage of UV
emission, the optical colors of both novae look quite red.

\begin{figure}
\includegraphics[width=0.5\textwidth]{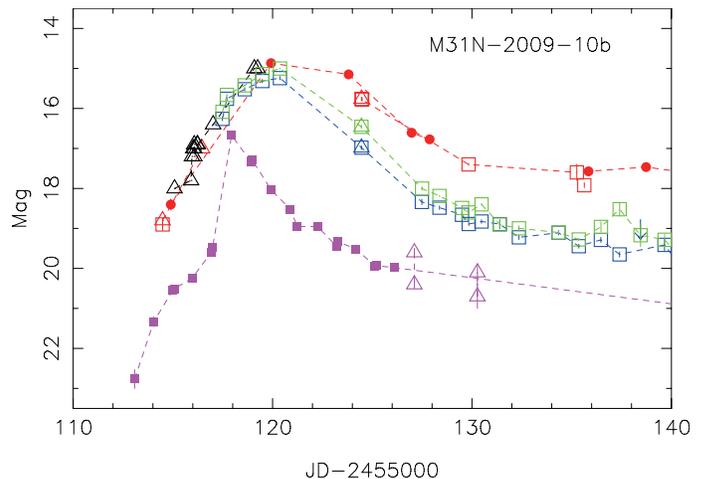}
\caption{The multiband light curve of M31 N2009-10b focusing on its early
evolution.  We can clearly see that the UV peaks prior to the peaks in B, V 
and R bands by about two days. When the UV decays, the optical part is quite
red. \label{2009-10b}}
\end{figure}

Of the two, 2009-10b is brighter, has better light curve coverage in PTF,
{\it GALEX} and other bands, and is thus a better case for further examination.
  Figure
\ref{2009-10b} zooms in on the early multi-band light curve. Table
\ref{2009-10b-table} lists the rise rate, maximum magnitudes, dates and
decline rates by $t_2$ and $t_3$ in the NUV, B, V and R bands. The NUV data
are from {\it GALEX}.  Part of R band data are from PTF. B, V and part of R
band data are published in \citet{SDH2011}. We can see clearly that the NUV
peak is two-to-three days earlier than the B, V and R peaks.  The rise rates in B, V
and R band are consistent while NUV emission rises swiftly. The decline
rates of NUV, B and V are similar while R band emission decays slower.

2010-11a is possibly the faintest nova in our sample with a peak observed
magnitude of R $\sim18.8$ (see Figure~\ref{lc6}). It was first detected by {\it
SWIFT} \citep{ATel3061} and then we reported its optical counterpart
\citep{ATel3066}.  The strong UV emission peaked several days prior to its
optical maximum. Near its optical maximum, we carried out a photometric
follow-up in i', r', g' and B bands with the Palomar 60-inch telescope and
observed that this nova had red optical colors, as shown in
Table~\ref{2010-11a}.

The UV observations and the color information suggest that UV and optical
emission originates from different radiative regions because they cannot be
simply interpreted as the thermal radiation of a receding spherical
photosphere which has a large optical depth in local thermal equilibrium.
Because UV emission is only generated deep inside the envelope, the result
may suggest that the envelope has holes from which we can see deep inside. 

The models of \citet{Shaviv2001,KH2005,KH2007} propose a reduced effective
opacity of the nova envelope to explain the super-Eddington phase,
resulting from a porous structure of the envelope caused by its fluid
instability.  We may use a similar scenario of asphericity of nova
outbursts to qualitatively explain the red UV novae. UV photons are
generated deep under the envelope at the surface of the WD. In the
envelope, the bottom is heated to very high temperatures by thermonuclear
runaway, and thus instabilities like Rayleigh-B$\acute{\rm e}$nard
convection are likely to occur. This instability causes the mass
distribution in the nova envelope to deviate from a spherical distribution.
Thus, along some lines of sight to the center of the outburst, the mass is
less and so is the opacity.  The UV photons then escape from these
directions and are observed by us if we are observing along one of these
lines of sight. The red optical emission still comes from the reprocessing
of the envelope in the other directions.  The ratio between the UV and
optical fluxes reflects the angular size of the cone. If the instability
produces an extreme feature with very low mass and very low opacity along
the line of sight, we can observe the central UV photons directly.  This
may be what we are observing in 2009-10b and 2010-11a.  Another possible
explanation could be very low abundances of chemical species that absorb in
the UV, although this may not explain the red optical colors.

The uniqueness of the UV-optical light curves of 2009-10b and 2010-11a suggest
either their super-Eddington phases or their chemical abundances (or both) are
very different from the majority of novae \citep{CAG2002}.  They may possibly
represent a population that deviates from any empirical relation derived from
large nova samples, such as MMRD relation.  In fact, 2009-10b is a luminous
outlier of the MMRD relation (see Figure~\ref{MMRD}).  Unfortunately, due to
its faintness, we were unable to measure the $t_2$ decline rate for the nova
2010-11a.

\begin{deluxetable}{cccc}
  \tablewidth{0pt}
  \tablecaption{P60 photometric follow-up of M31N 2010-11a\label{2010-11a}}
  \tablehead{\colhead{JD-2455500(days)} & \colhead{Bands} & \colhead{Magnitude} & \colhead{Magnitude Error}}
  \startdata
  34.817 & i' & 18.80 & 0.10 \\
  34.819 & i' & 18.84 & 0.09 \\
  34.822 & i' & 18.89 & 0.09 \\
  34.824 & r' & 18.91 & 0.08 \\
  34.826 & r' & 19.04 & 0.09 \\
  34.838 & r' & 18.96 & 0.08 \\
  34.831 & r' & 18.89 & 0.09 \\
  34.833 & r' & 19.04 & 0.11 \\
  34.835 & g' & 19.20 & 0.09 \\
  34.838 & g' & 19.32 & 0.11 \\
  34.840 & g' & 19.12 & 0.09 \\
  34.842 & B  & 19.40 & 0.13 \\
  34.844 & B  & 19.37 & 0.16 \\
  34.847 & B  & 19.43 & 0.15
  \enddata
\end{deluxetable}

\begin{deluxetable}{ccccc}
  \tablewidth{0pt}
  \tablecaption{2009-10b NUV and optical properties \label{2009-10b-table}}
  \tablehead{}
  \startdata
  Bands & NUV\tablenotemark{1} & B\tablenotemark{2} & V\tablenotemark{2} & R\tablenotemark{3} \\
  rise rate (mag/day) & 1.2 & 0.4 & 0.4 & 0.7  \\
  peak magnitude & 16.68  & 15.24 & 15.00 & 14.84 \\
  peak date (JD-2455110)  & 7.95 & 10.37 & 10.37 & 9.95 \\
  $t_2$(days) & 3 & 5 & 5 & 8 \\
  $t_3$(days) & 7 & 7 & 7 & 17
  \enddata
  \tablenotetext{1}{{\it GALEX} observations;}
  \tablenotetext{2}{data from \citet{SDH2011};}
  \tablenotetext{3}{data from PTF and \citet{SDH2011}.}
\end{deluxetable}

\section{MMRD Revisited}
\label{section_MMRD}
Novae exhibit a relation between the maximum magnitude and the rate of
decline \citep[MMRD][]{dL1995, KCK2010}. We can assess this relation
with our nova sample in M31. 

One of main issues in evaluating the MMRD relation is estimating the dust
extinction.  A multitude of methods using color and Balmer decrement have
been used in literature to estimate the extinction \citep{DBK2006,
Kogure1961, KCK2010}.  The PTF nova observations, however, provide no color
information, being done only in the R band.  Thus, we apply a mean foreground 
dust extinction.  The foreground
extinction of M31, $E(B-V)=0.062$mag \citep{SFD1998}, leads to
\begin{eqnarray}
	A_g=3.793\times E(B-V)=0.234.
\end{eqnarray}
For a better comparison with previous researches
\citep[e.g.,][]{KCK2010}, following Shafter et al.(2009), we use the colors
of an A5V star ($T=8200K$) to transform $R$ magnitudes to $g$ magnitudes
\citep{JGA2006}. 

\begin{figure}
\includegraphics[width=0.5\textwidth]{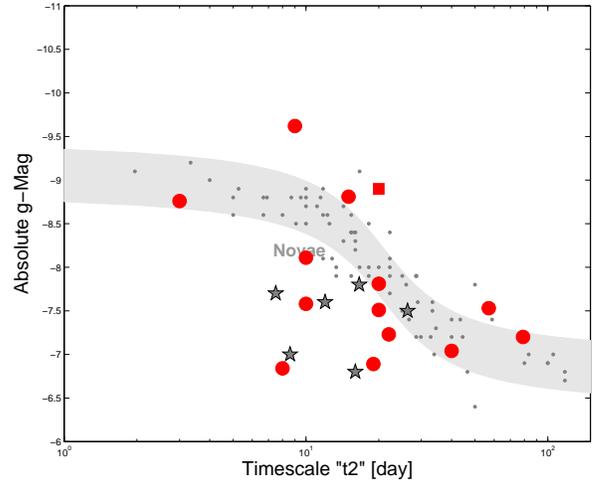}
\caption{The empirical relation for novae between the maximum magnitude and
the rate of decline (MMRD).  The gray region denotes the MMRD from
\citet{dL1995}, while the small dark gray solid dots represents their nova
sample. The six gray stars show the sample from \citet{KCK2010}.  Red circles 
and a red rectangle are the M31 novae from this paper.  The red rectangle 
denotes the bright recurrent nova M31N\,2010-12a.
\label{MMRD}}
\end{figure}

The MMRD relation is presented as the gray band in Figure \ref{MMRD}.  We
overplot on this diagram our M31 novae (red points) and compare them to the
novae from \citet{dL1995} (gray points) and \citet{KCK2010} (gray stars).
Without a local extinction correction, most of our peak magnitudes are
underestimated.  In \citet{KCK2010}, the total extinction correction for
M31 novae is roughly $A_g=0.24$ except that two of them ($20\%$) have a
correction of $A_g>1.0$.  It is possible that local extinction for most
novae in our sample is small. 

In Figure \ref{MMRD}, we can see five significant outliers from the MMRD
region in our sample.  The brightest outlier is M31N\,2010-10b (see
discussion in Section \ref{UV2009-10b}).  The other luminous outlier is PT
And ($=$M31N\,2010-12a). M31N\,2009-12a, M31N\,2010-06b and M31N\,2010-06c
lie in the faint and fast-decline rate zone below the MMRD relation
shown by the gray band.  \citet{KCK2010} also found several novae in this
zone.  \citet{KCK2010} proposed the possibility that the outliers may be
RNe. However, none of the three faint outliers in our sample are 
known to be recurrent.
The remaining novae reside roughly along the MMRD region. 

\section{Conclusion}
\label{conclusion}

In this paper, we tabulate the photometric measurements of twenty-nine 
classical novae in M31
from PTF data and present their optical light curves, eight of which also
have joint UV-optical light curves from both PTF and {\it GALEX}. The main
findings are as follows:
\begin{enumerate}
  \item We detected three recurrent novae. They show similar properties to 
        recurrent novae 
	in the Galaxy \citep{Schaefer2010}. We obtained a spectrum of the 
	recent outburst of PT And (also known as M31N\,2010-12a). While 
	previously identified as a dwarf nova in the Milky Way \citep{AZ2000},
	we find the blueshifts in our spectrum to be consistent with
	the event being a recurrent nova in M31. The light curve of 2010-12a
	resembles those of previous outbursts, suggesting the same physical
	conditions in the recurrent outbursts. Our search at all other 
	historical nova positions finds no additional outburst candidates.
	From this we derive a recurrent nova rate of about three per year
	in M31.
  \item The rise behaviors of M31 classical novae are diverse and we classify them
	morphologically into S (smooth) and J (jittering) classes.  We find
	that the S-class novae rise significantly faster than the J-class
	novae. We see no dependence between rise-time and maximum magnitude.
  \item During the ten minute cadence PTF observations the rising light
	curve of two novae were recorded with unprecedented detail.  These
	novae showed smooth and consistent rising light curves without large amplitude 
	variability. 
  \item The declining light curves are also divided morphologically into S 
	and J classes.  The relative population of different classes in
	M31 is roughly consistent with that in the Galaxy.
  \item Three quarters of the well-sampled novae obey the empirical 
	MMRD relation. The remaining one quarter (five events)
	are consistent with the ``faint and fast'' outliers 
	found by \citet{KCK2010}.
  \item Some novae have UV detections prior to or at the same time as the
	first optical detection.  Moreover, 2009-10b and 2010-11a have UV
	peaks prior to optical maxima while both are quite red in the UV
	decay phases.  These observations are inconsistent with
	theoretical predictions of delayed UV emission in the iron curtain 
	stage, possibly suggesting extreme density or abundance variations
	for these novae.
  \item Apart from 2010-10f in Bol~126, we find no other novae in all the
	cataloged M31 GCs.  We derive a GC nova rate in M31 of 
	$\sim\,1$ yr$^{-1}$.  This nova rate is not enhanced relative to 
	the rate in M31 outside of the GC system. 
\end{enumerate}

We realize the novae reported here are only a subset of all the classical novae in
M31 as the number is smaller than expected \citep[e.g., 65~nova~yr$^{-1}$]
[]{DBK2006}. We did not undertake a systematic search for classical novae in
PTF images during the 2009 and 2010 seasons. Since the FoV and survey depth
of PTF should be sensitive to all classical novae (weather-permitting), we have started
a more complete search for classical novae in M31 in the 2011 season. 
For 2012, we intend to run a real-time search for novae.  Our
photometric and spectroscopic follow-up will enable us to better understand
their properties and to estimate the local extinction correction. 

Besides optical bands, future UV surveys are warranted for nova studies.
Some novae peak in the UV before the optical. Such novae could 
provide important insights into the development 
and structure of the nova wind.  It is also important to know if novae like
M31N\,2009-10b and M31N\,2010-11a are rare or common and if they are always
outliers in the MMRD diagram.

Ongoing and upcoming synoptic surveys (e.g., Palomar Transient Factory and 
Next Generation Transient Facilities) will continue
to monitor the Andromeda Galaxy. The long time baseline will give us
a larger sample of recurrent novae.
As noted earlier, recurrent novae could be the progenitors of Ia 
supernovae. Thus the increased sample and timely follow-up will give us a 
comprehensive understanding of rate of recurrent 
novae--an important clue to their endpoint. 

\acknowledgments
We thank Marina Orio and Sumin Tang for valuable discussions. We thank 
the Weizmann Monitoring Team (A. Gal-Yam, I. Arcavi, D. Polishook, 
A. Sternberg, O. Yaron, D. Xu) for daily monitoring of transient candidates
from the PTF discovery stream.

MMK acknowledges support from the Hubble Fellowship and the 
Carnegie-Princeton Fellowship. SBC wishes to acknowledge generous 
support from Gary and Cynthia
Bengier, the Richard and Rhoda Goldman Fund, National Aeronautics and
Space Administration (NASA)/{\it Swift} grant NNX10AI21G, NASA/{\it
  Fermi} grant NNX1OA057G, and National Science Foundation (NSF) grant
AST--0908886. 

This research was supported in part by Tsinghua Center for Astrophysics
(THCA), by the National Natural Science Foundation of China (NSFC) grants
10373009, 10533020 and 11073014 at Tsinghua University, by the Ministry of Science
and Technology (MOST) under State Key Development Program for Basic Research
grant 2012CB821800, by the Tsinghua University Initiative Scientific Research
Program, by the SRFDP 20050003088, 200800030071 and 20110002110008, and by the Yangtze
Endowment and the SRFDP 20050003088 and 200800030071 from the Ministry of
Education at Tsinghua University. 

GALEX (Galaxy Evolution Explorer) is a NASA Small Explorer, launched in
2003 April. We gratefully acknowledge NASA's support for construction,
operation, and science analysis for the GALEX mission, developed in
cooperation with the Centre National d'Etudes Spatiales of France and the
Korean Ministry of Science and Technology.

The National Energy Research Scientific Computing Center, which is supported
by the Office of Science of the U.S. Department of Energy under Contract
No. DE-AC02-05CH11231, provided staff, computational resources and data
storage for PTF. 

This research has made use of the NASA/IPAC Extragalactic Database (NED)
which is operated by the Jet Propulsion Laboratory, California Institute of
Technology, under contract with the National Aeronautics and Space
Administration.

\bibliographystyle{apj}
\bibliography{refs,ATel,CBET}

\end{document}